\documentclass[aps,prd,superscriptaddress,nofootinbib,eqsecnum,twocolumn]{revtex4}

\pdfoutput=1

\usepackage{amsfonts}
\usepackage{amsmath}
\usepackage{amssymb}
\usepackage{graphicx,color}
\usepackage{float}
\usepackage{hyperref}
\usepackage{subfigure}
\usepackage{mathtools,slashed}
\usepackage{ulem}
\usepackage{lipsum}
 
\def\be {\begin{equation}}
\def\ee {\end{equation}}
\def\bea {\begin{eqnarray}}
\def\eea {\end{eqnarray}}
\def\bc {\begin{center}}
\def\ec {\end{center}}

\def\mn {\mu\nu}

\def\nn {\nonumber}
\newcommand \Tr{\operatorname{\text{Tr}}}
\def\sp {\shortparallel}

\date{\today}

\begin{document}
\title{Heavy quark diffusion coefficients in magnetized quark-gluon plasma} 
\author{Aritra Bandyopadhyay}
\affiliation{Institut für Theoretische Physik, Universität Heidelberg, Philosophenweg 16, 69120 Heidelberg, Germany}

\begin{abstract}{
We evaluate the heavy quark momentum diffusion coefficients in a hot magnetized medium for the most general scenario of any arbitrary values of the external magnetic field. We choose to work with the systematic way of incorporating the effect of the magnetic field, by using the effective gluon and quark propagators, generalized for a hot and magnetized medium. To get gauge independent analytic form factors valid through all Landau levels, we apply the Hard Thermal Loop (HTL) technique for the resummed effective gluon propagator. The derived effective HTL gluon propagator and the generalized version of Schwinger quark propagator subsequently allow us to analytically evaluate the longitudinal and transverse momentum diffusion coefficients for charm and bottom quarks beyond the static limit. Within the static limit we also explore another way of incorporating the effect of the magnetic field, i.e. through the magnetized medium modified Debye mass and compare the results to justify the need for structural changes. }
\end{abstract}

\maketitle
\newpage

\section{Introduction}
\label{sec1}

Heavy quarks have been extensively studied within the heavy-ion community as a significant hard probe for characterizing the properties of hot and dense quark matter~\cite{Thoma:1990fm,Braaten:1991jj,Braaten:1991we,Mustafa:2004dr,Moore:2004tg,Casalderrey-Solana:2006fio,CaronHuot:2007gq,
Casalderrey-Solana:2007ahi,CaronHuot:2008uh,Gossiaux:2008jv,Peigne:2008nd,Beraudo:2009pe,Monteno:2011gq,Alberico:2011zy,Alberico:2013bza,
Rapp:2018qla,Singh:2018wps,He:2022ywp,Madni:2022bea,Debnath:2023dhs,Altenkort:2023oms}. Considering the non-centrality of the HIC experiments and the subsequent generation of strong magnetic fields~\cite{Skokov:2009qp,Deng:2012pc,Bloczynski:2012en,Tuchin:2014iua,bzdak,McLerran}, studies related to heavy quarks have recently been extended to magnetized medium~\cite{Fukushima:2015wck,Sadofyev:2015tmb,Finazzo:2016mhm,Kurian:2019nna, Singh:2020faa,Singh:2020fsj,Bandyopadhyay:2021zlm,Mazumder:2022jjo,Ghosh:2022sxi,Nilima:2022tmz,Satapathy:2022xdw,Jamal:2023ncn,Dey:2023lco}. Most of the heavy quark (HQ) dynamical studies within magnetized medium have been restricted to limiting scenarios, i.e. adhering to strong or weak field approximations. In the present work we go beyond those limiting scenarios to tackle the most general case of arbitrary valued external magnetic fields, for the first time in literature. In the following, we further enlarge on our motivation as a prelude to the proceeding calculation. 

HQ momentum diffusion coefficients are the essential theoretical inputs required to describe the HQ evolution using Langevin equations~\cite{Beraudo:2009pe,Monteno:2011gq,Alberico:2011zy,Alberico:2013bza,Bu:2022oty}. This widely adopted approach assumes {\it external} HQ receiving random {\it kicks} from the thermal partons in the bulk medium. The HQ diffusion coefficients, along with the drag coefficient acutely influence the phenomenology relevant to HQs, thereby affecting the corresponding theoretical predictions for the relevant experimental observable~\cite{Rapp:2018qla}. While evaluating these parameters, numerous studies typically consider the non-relativistic static limit of the HQ, a reasonable approximation for low-momentum charm and bottom quarks~\cite{CaronHuot:2007gq,Fukushima:2015wck}. The need for the more general relativistic dynamic limit of the HQ comes from the current HIC experiments for heavy flavor sector spanning well into the high momentum region~\cite{Moore:2004tg,Beraudo:2009pe,Bandyopadhyay:2021zlm}. 

In absence of the magnetic field and within the static limit (i.e. $p\approx 0$, $M\gg T$, $p$ and $M$ being the HQ momentum and mass respectively) there is no anisotropy imposed on the system. Hence we have a single diffusion coefficient $\kappa$, resulting in mean squared HQ momentum transfer per unit time to be $3\kappa$. This $3\kappa$ can be evaluated considering the scattering processes of thermally populated light quarks and gluons with the HQ, i.e. $2\leftrightarrow 2$ scattering of $qH\leftrightarrow qH$ and $gH\leftrightarrow gH$ ($q \rightarrow$ quark, $g\rightarrow$ gluon and $H\rightarrow$ HQ). Because of the large mass difference and relatively small energy transfer, the $t$-channel scatterings mediated by gluons dominate these processes at leading order in the strong coupling and the scattering particles can be considered as quasiparticles within the thermally equilibrated matter. Several studies have evaluated the diffusion coefficient $\kappa$ using various techniques producing interesting results, e.g. perturbative results up to NLO~\cite{CaronHuot:2007gq}, within Gribov-Zwanziger action~\cite{Madni:2022bea} and the very recent lattice QCD evaluation~\cite{Altenkort:2023oms}. 

Going beyond the static limit associates a finite velocity $\gamma v \lesssim 1$ (i.e. $p=\gamma M v \lesssim M$) with the HQ, subsequently introducing an anisotropy in the system generated from the movement of HQ in a preferred direction. Hence $\kappa$ breaks down into longitudinal and transverse parts, i.e. $3\kappa \to  \kappa_L + 2\kappa_T$. There also have been several beyond the static limit perturbative calculations for the HQ diffusion coefficients which have usually incorporated the  Hard Thermal Loop (HTL) resummation method for the hot medium~\cite{Braaten:1991jj,Braaten:1991we,Thoma:1990fm,Moore:2004tg,CaronHuot:2007gq,Beraudo:2009pe,Monteno:2011gq,Alberico:2011zy,Alberico:2013bza}. 
 
The presence of an external magnetic field brings another new anisotropy into the system along with some interesting new questions regarding the incorporation of an extra scale $eB$. HQ diffusion coefficients in a hot and magnetized medium are presently being explored both within and beyond the static limit~\cite{Fukushima:2015wck,Sadofyev:2015tmb,Finazzo:2016mhm,Kurian:2019nna,Singh:2020faa,Singh:2020fsj,Bandyopadhyay:2021zlm, Mazumder:2022jjo}, most of which have been done considering the Lowest-Landau-Level (LLL) approximation, i.e. assuming $eB \gg T^2$~\cite{Fukushima:2015wck,Singh:2020faa,Singh:2020fsj,Bandyopadhyay:2021zlm} or the weak magnetic field approximation~\cite{Dey:2023lco}. The validity of the LLL or weak field approximations can be argued assuming the strength of the magnetic field generated in non-central HICs and its time dependence. But extending the approximated calculations to the most general scenario of arbitrary external magnetic fields puts an end to all those arguments. It also unburdens us from extra constraints imposed on the magnetic field scale $eB$. The evaluation of the HQ momentum diffusion coefficients (which are directly dependent on the HQ scattering rate) in presence of arbitrary valued magnetic fields requires us to compute the effective gluon propagator generalized for hot magnetized medium beyond the LLL approximation. On this front we employ HTL approximations to calculate the form factors of the effective generalized gluon propagator valid across all Landau levels. An alternate way of incorporating the effect of the magnetic field has been utilized recently in the calculations of HQ potential and HQ energy loss where the sole medium effect has been assumed to be channeled through the medium dependent Debye mass~\cite{Ghosh:2022sxi,Nilima:2022tmz,Jamal:2023ncn}. Within the static limit, we will compare the results originating from both these methods and discuss the limitations of using the apparently crude approximation of medium modified Debye mass.

In this paper, we aim to address an important state of the art problem, i.e. the  calculation of the heavy quark momentum diffusion coefficients beyond the static limit in a quark-gluon plasma under the influence of an arbitrary external magnetic field. To explore the same :
\begin{enumerate}
    \item We consider a HQ moving with a velocity $\vec{v}$ in presence of an anisotropic $\vec{B} = B \hat{z}$ and analytically derive the full results for the longitudinal and transverse momentum diffusion coefficients for charm and bottom quarks.  
    \item  Though the heavy quark mass $M\gg \sqrt{eB},T$ is considered to be the largest scale of the system, unlike LLL or weak field approximations we do not restrict ourselves with further scale hierarchies with respect to $eB$ and $T$. 
    \item Similar to Refs~\cite{Fukushima:2015wck,Bandyopadhyay:2021zlm}, here we also work within the HTL approximation with a further constraint $\alpha_s eB \ll T^2$ ($\alpha_s$ being the QCD running coupling), which helps us neglect the soft self energy corrections of the quarks and gluons while evaluating the scattering rate. 
\end{enumerate}

The rest of this paper is organized as follows. In the following section (section \ref{sec2}) we provide the formalism and summarize the calculational steps, to be carried out in this work. In section \ref{sec3} we evaluate the expressions for various form factors required to construct the most general one-loop effective gluon propagator in a hot magnetized medium. Section~\ref{sec4} witnesses the computation of the HQ scattering rate in an arbitrarily magnetized medium beyond the static limit. In section \ref{sec5} we provide the final expressions for the momentum diffusion coefficients of HQ in a magnetized medium for both $\vec{v} \sp \vec{B}$ and $\vec{v}\perp\vec{B}$. We divide the results into two sections. Section \ref{sec6} contains special discussions about the HQ momentum diffusion coefficients within the static limit of HQ and comparison between two alternate ways of incorporating the magnetic field induced effects within them. In section \ref{sec7} we show our estimations of the HQ momentum diffusion coefficients beyond the HQ static limit and discuss the results. Finally we summarize and conclude in section \ref{sec8}. Novelty of the present work is well reflected in the explicit calculations presented in the Appendices~\ref{appA},\ref{appB},\ref{appC},\ref{appD}, and \ref{appE}.


\section{Formalism}
\label{sec2}

The present work evaluates the HQ diffusion coefficients in the general most scenario of hot and magnetized medium, also assuming the HQ to be relativistic (i.e. going beyond the static limit). Before going into the details of the evaluation procedure, in this section we will give an overview of the same and clarify the notations to be used throughout the paper.

When a HQ traverses through the medium, it encounters collisions with other partons. Because the HQ has a much higher energy scale than the temperature of the medium, i.e. $P\equiv (M, p) \equiv (E, v) \gg T$, it usually takes a large number of collisions (around $M/T$ within and $p/T$ beyond the static limit of HQ) to change the HQ momentum by a substantial amount. Hence one can approximate the interaction of the HQ with the medium in a simplified way, such that it becomes a series of uncorrelated momentum kicks. At $T\neq 0$, these uncorrelated momentum kicks can be comprehended as originating from the scatterings faced by the HQ with the thermally populated light quarks and gluons. As a result of that, the transport coefficients, i.e. in our case the momentum diffusion coefficients ($\kappa$) in our case, are directly related to the corresponding scattering/interaction rate ($\Gamma$), as is shown below through explicit expressions.

In the presence of an external magnetic field and considering that the HQ is moving in a particular direction, i.e. going beyond the static limit of HQ, the complex interplay between anisotropies generated from the preferred directions of HQ momentum/velocity and the magnetic field leads to nontrivial scenarios. In this situation it is useful to work with two simple cases, i.e. $\vec{v} \sp \vec{B}$ and $\vec{v} \perp \vec{B}$. The first case $\vec{v} \sp \vec{B}$ leads to two different diffusion coefficients $\kappa_L$ and $\kappa_T$, related to the HQ scattering rate as
 \begin{align}
\kappa_T (p) = \frac{1}{2}\int d^3q\frac{d~\Gamma(v)}{d^3q}q_\perp^2,~~ \kappa_L (p) = \int d^3q\frac{d~\Gamma(v)}{d^3q}q_z^2.
\label{coeffs_case1}
\end{align}
On the other hand, $\vec{v} \perp \vec{B}$ case generates three different diffusion coefficients $\kappa_j$'s ($j=\{x,y,z\}\equiv\{1,2,3\}$), i.e.
\begin{align}
\kappa_j (p) = \int d^3q\frac{d~\Gamma(v)}{d^3q}q_j^2.
\label{coeffs_case2}
\end{align}
It is straightforward to realize that retracing the static limit ($\vec{v}\rightarrow 0$) within a magnetized medium would mean there is only one anisotropy given by the specific direction of $\vec{B}$ and would cause the  $\vec{v} \perp \vec{B}$ case to vanish.

\begin{figure}
\begin{center}
\includegraphics[scale=0.4]{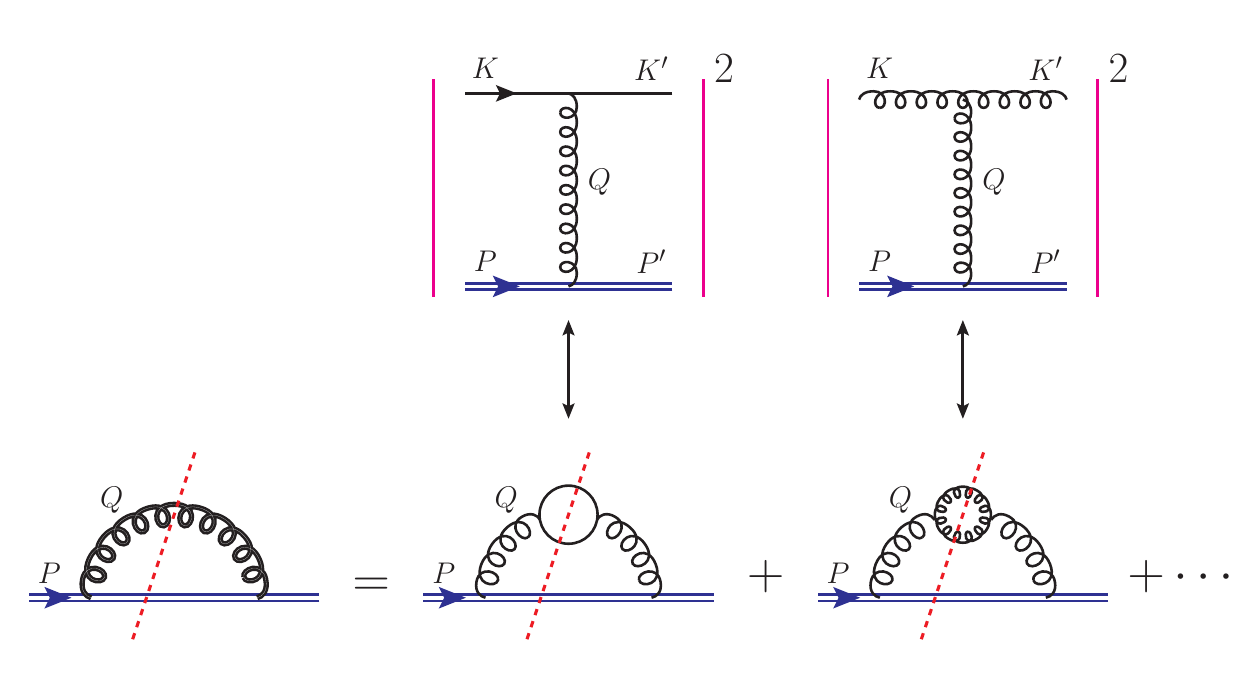}
\caption{The equivalences of the $t$-channel scatterings of heavy quarks due to thermally generated light quarks and gluons, $qH\rightarrow qH$ (left) and $gH\rightarrow gH$ (right) are shown, as they can also be expressed as the cut (imaginary) part of the HQ self energy. An HTL resummed heavy quark self-energy with effective gluon propagator takes into account the diagrams for the hard process among others.} 
\label{hq_sqme_alt}
\end{center}
\end{figure}

From Eqs.\ref{coeffs_case1} and \ref{coeffs_case2} one can easily deduct that to evaluate the momentum diffusion coefficients, we first need to evaluate the scattering rate or interaction rate of the $2\leftrightarrow 2$ scatterings between the light quark/gluon and the HQ (i.e. $qH\leftrightarrow qH$ and $gH\leftrightarrow gH$). The dominant $t$-channel scattering processes involving them are diagrammatically portrayed in the top half of Fig.~\ref{hq_sqme_alt}. Now, to evaluate the corresponding scattering rates, we use an effective approach, first provided by Weldon~\cite{Weldon:1983jn}. In this approach we can express the t-channel $2\leftrightarrow 2$ scatterings involving HQ as cut/imaginary parts of the HQ self energy, demonstrated in Fig.~\ref{hq_sqme_alt}. Using this technique the expression for the scattering rate ($\Gamma(P)$) comes out to be 
\begin{align}
\Gamma(P) = -\frac{1}{2E}~\frac{1}{1+e^{-E/T}}~\Tr\left[(\slashed{P}+M)~{\rm Im}~\Sigma(p_0+i\epsilon,{\vec{p}})\right],
\label{interaction_rate2}
\end{align}
where $\Sigma(P)$'s represent the two-loop HQ self energy diagrams involving second-order quark and gluon loops. The hard contribution of the scattering rate comes from cutting these $\Sigma(P)$'s, depicted by the right hand side of the diagrammatic equation shown in the bottom half of Fig.~\ref{hq_sqme_alt}. For the present study we also want to incorporate the soft contributions, i.e. where the momentum flowing through the mediating gluon ($Q$) is considered to be soft. In this scenario, the hard thermal loop (HTL) corrections to the gluon propagator contribute at the leading order in the strong coupling constant, which in turn suggests that resummation must be taken into account. At this point we will emphasize the advantage of using Eq.~\ref{interaction_rate2}, which allows us to include all the necessary resummations by applying the imaginary time formalism of the thermal field theory. Hence effectively to include the resummation, all we have to do is to replace the several separate $\Sigma(P)$'s by a sole effective HQ self energy (which we will also symbolize as $\Sigma(P)$ from this point onward) with an HTL resummed effective gluon propagator (depicted by the left hand side of the diagrammatic equation shown in the bottom half of Fig.~\ref{hq_sqme_alt}), which in a magnetized medium can be expressed as, 
\bea
\Sigma(P) = ig^2\int\frac{d^4Q}{(2\pi)^4}\mathcal{D}^{\mn}(Q)\gamma_\mu S_m(P-Q)\gamma_\nu.
\label{htl_eff_self_energy}
\eea
Here $\mathcal{D}^{\mn}(Q)$ is the effective gluon propagator and $S_m(P-Q)$ is the heavy fermion propagator in presence of an external magnetic field.

Next we will discuss the individual components of Eq.~\eqref{htl_eff_self_energy} in details. First of all, the heavy fermion propagator $S_m(P-Q\equiv K)$ in presence of an arbitrary external magnetic field is given by~\cite{Schwinger:1951nm,Gusynin:1995nb,Calucci:1993fi}, 
\begin{eqnarray}
S_m(K) = e^{-\frac{k_\perp^2}{|q_fB|}} \sum_{l=0}^{\infty}
\frac{(-1)^l D_l(q_fB, K)}{K_\shortparallel^2-M^2-2lq_fB},
\label{decomposed_propagator}
\end{eqnarray}
with $l=0,\, 1,\, 2, \ldots$, denoting the Landau levels  and  
\begin{align}
& D_l(q_fB,K) =
(\slashed{K}_\shortparallel+M)\Bigl((1-i\gamma^1\gamma^2)L_l\left(\xi_k^\perp\right)
\nonumber\\ 
&~~ -(1+i\gamma^1\gamma^2)L_{l-1}\left(\xi_k^\perp\right)\Bigr)-
4\slashed{k}_\perp L_{l-1}^1\left(\xi_k^\perp\right),
\label{d_l}
\end{align}
where $\xi_k^\perp = \frac{2k_\perp^2}{q_fB}$ and $L_l^\alpha (\xi_k^\perp)$ is the generalized Laguerre polynomial, defined as
\begin{eqnarray}
(1-z)^{-(\alpha+1)}\exp\left(\frac{z~\xi_k^\perp}{z-1}\right) =
  \sum_{l=0}^{\infty} L_l^\alpha(\xi_k^\perp) z^l.
\end{eqnarray}
where $q_f$ is the fermionic charge for a particular flavor $f$. $K^\mu\equiv (K_\sp^\mu,k_\perp^\mu)$ is the fermionic four momentum with $K_\shortparallel^\mu = (k^0,0,0,k^z)$ and $k_\perp^\mu = (0,k^x,k^y,0)$. The metric tensor can also be broken down as $g^{\mu\nu} = g_\shortparallel^{\mu\nu} + g_\perp^{\mu\nu}$, with $g_\shortparallel^{\mu\nu} = \textsf{diag}(1,0,0,-1)$ and $g_\perp^{\mu\nu} = \textsf{diag}(0,-1,-1,0)$, such that the individual components satisfy $K^2=K_\sp^2-k_\perp^2$, i.e. $K_\sp^2 =k_0^2-k_z^2$ and $k_\perp^2 = k_x^2+k_y^2$. 

Next we focus on the effective gluon propagator. Several recent advances have been made on the general structures of the fermion and gauge boson self-energies with propagators at finite temperature and in presence of an external magnetic field~\cite{Shabad:2010hx,Hattori:2012je,Bordag:2008wp,Chao:2014wla,Mueller:2014tea,Das:2017vfh,Ayala:2018ina,Karmakar:2018aig,Ayala:2020wzl,Ayala:2021lor} as well as thermo-magnetic correction to the quark-gluon vertex~\cite{Ayala:2014uua,Haque:2017nxq}. Out of these choices we usually work with the effective gluon propagator in a hot and magnetized medium from Ref.~\cite{Karmakar:2018aig}, i.e.,
\begin{align}
&\mathcal{D}^{\mn}(Q)=\frac{\xi Q^{\mu}Q^{\nu}}{Q^4}+\sum_{i=1}^4\mathcal{J}_i\Delta_i^{\mn},
\label{gauge_prop}
\end{align}
where $\xi$ is the gauge parameter, $\Delta_i^{\mn}$'s are the constructed tensor basis and $\mathcal{J}_i$'s are the corresponding coefficients. Various $\Delta_i^{\mn}$'s can be expressed as 
\begin{subequations}
\begin{align}
\Delta_1^{\mu\nu} &= \frac{1}{\bar{u}^2} \bar{u}^\mu\bar{u}^\nu, \label{D1munu}\\
\Delta_2^{\mu\nu}  &=g_{\perp}^{\mn}-\frac{Q^{\mu}_{\perp}Q^{\nu}_{\perp}}{Q_{\perp}^2}, \label{D2munu}\\
\Delta_3^{\mu\nu}  &=  \frac{{\bar n}^\mu {\bar n}^\nu}{\bar n^2}, \label{D3munu}\\
\Delta_4^{\mu\nu} &= \frac{\bar u^{\mu}\bar n^{\nu}+\bar u^{\nu}\bar n^{\mu}}{\sqrt{\bar u^2}\sqrt{\bar n^2}},\label{D4munu}
\end{align}
\label{Deltamunu}
\end{subequations}
with 
\begin{subequations}
\begin{align}
\bar{u}^\mu &= u^\mu - \frac{q_0 Q^\mu}{Q^2},\\
g_\perp^{\mu\nu} &= \textsf{diag}(0,-1,-1,0),\\
Q_\perp^\mu Q^\perp_{\mu} &=Q_\perp^2= Q^2-Q_\sp^2=-q_\perp^2,\\
\bar{n}^\mu &= n^\mu - \frac{q_3 Q^\mu}{q^2}+ \frac{q_0q_3 u^\mu}{q^2},
\end{align}
\end{subequations}
where $u^\mu =(1,0,0,0)$ is the heat bath velocity and $n_\mu = (0,0,0,1)$ is defined uniquely as the projection of the electromagnetic field tensor $F_{\mn}$ along $u^\mu$. 

\begin{figure}
\centering\includegraphics[scale=0.42]{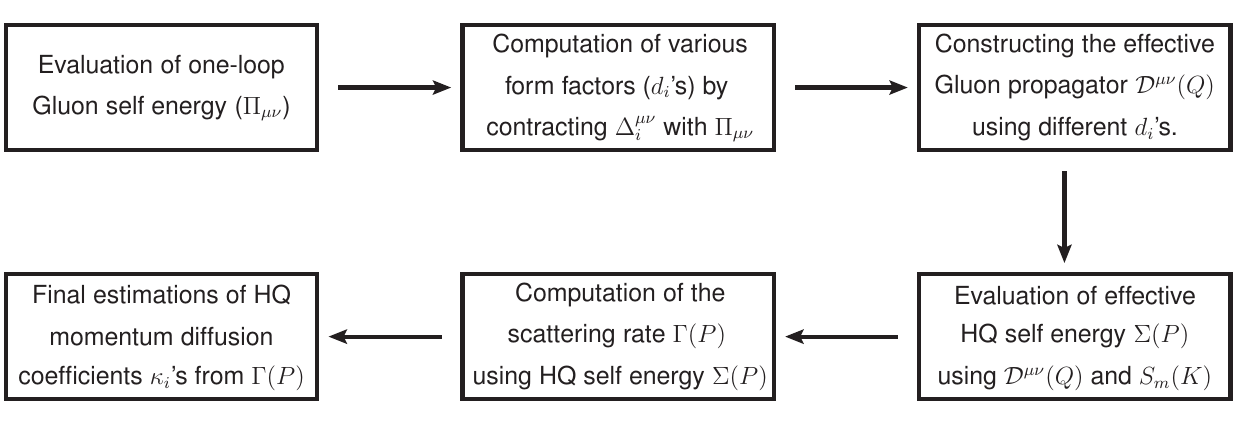}
\caption{A flow chart summarizing the calculational steps to be followed from section \ref{sec3}.}
\label{flowchart}
\end{figure}

Subsequently corresponding coefficients $\mathcal{J}_i$'s are given as
\begin{subequations}
\begin{align}
\mathcal{J}_1 &= \frac{(Q^2-d_3)}{(Q^2-d_1)(Q^2-d_3)-d_4^2}, \\
\mathcal{J}_2 &= \frac{1}{(Q^2-d_2)}, \\
\mathcal{J}_3 &=\frac{(Q^2-d_1)}{(Q^2-d_1)(Q^2-d_3)-d_4^2}, \\
\mathcal{J}_4 &= \frac{d_4}{(Q^2-d_1)(Q^2-d_3)-d_4^2},
\end{align}
\label{mathcaljs}
\end{subequations} 
where $d_i$'s are the form factors defined as 
\begin{subequations}
\begin{align}
d_1(Q) &= \Delta_1^{\mn}\Pi_{\mn}(Q), \label{ff_d1} \\
d_2(Q) &= \Delta_2^{\mn}\Pi_{\mn}(Q), \label{ff_d2} \\
d_3(Q) &= \Delta_3^{\mn}\Pi_{\mn}(Q), \label{ff_d3} \\
d_4(Q) &= \frac{1}{2}\Delta_4^{\mn}\Pi_{\mn}(Q) \label{ff_d4},
\end{align}
\label{ff_di}
\end{subequations} 
$\Pi_{\mn} (Q)$ being the one-loop gluon self energy. The form factors $d_i$'s have previously calculated only in the strong magnetic field limit within the lowest Landau level approximation in Ref.~\cite{Karmakar:2018aig} and subsequently used in Ref.~\cite{Bandyopadhyay:2021zlm} to explore the HQ dynamics. For the general scenario of any arbitrary external magnetic field, explicit evaluation of the form factors $d_i$'s requires the computation of one loop HTL gluon self energy including quarks residing in any arbitrary Landau levels, which we will discuss in the next section. Before jumping into that, readers should look at the calculational steps to be followed to obtain the final expressions for the HQ momentum diffusion coefficients, in the form of a flow chart (Fig.~\ref{flowchart}).

\section{Evaluation of the form factors from the one-loop HTL gluon self-energy}
\label{sec3}

In the previous section, we covered the basic formalism and summarized the calculational steps to be followed. As the first step of the same, in this section we will first compute the One loop gluon self energy ($\Pi_{\mn}$) in a hot and arbitrarily magnetized medium, which will then help us to evaluate various form factors ($d_i$'s) required to construct the HTL effective gluon propagator ($\mathcal{D}^{\mn}$). 

One loop gluon self energy in a hot and magnetized medium can be written as a combination of the pure glue Yang-Mills contribution $\Pi^g_{\mn}$ and the fermionic loop contribution $\Pi^m_{\mn}$, i.e. $\Pi_{\mn} = \Pi_{\mn}^m +\Pi_{\mn}^g$. The pure glue part $\Pi_{\mn}^{g}$ is unaffected in presence of magnetic field and can be written as
\bea
\Pi_{\mn}^{g}(Q)=-\frac{N_cg^2T^2}{3} \int \frac{d \Omega}{4 \pi}\left(\frac{q_{0} \hat{K}_{\mu} \hat{K}_{\nu}}{\hat{K} \cdot Q}-g_{\mu 0} g_{\nu 0}\right).
\label{Pi_mn_g}
\eea
As later we will be using it, at this point we also define the angular factor as $\mathcal{T}_Q$, given as 
\begin{align}
&\mathcal{T}_Q = \int \frac{d \Omega}{4 \pi}\frac{q_0}{\hat{K} \cdot Q} = \frac{q_0}{2q}\ln\left(\frac{q_0+q}{q_0-q} \right).
\end{align}

On the other hand the fermionic part of the one-loop gluon self-energy can be computed as
\begin{align}
\Pi_{\mu\nu}^m(Q) &= \sum_{f}\frac{ig^2}{2}\int\frac{d^4K}{(2\pi)^4}\textsf{Tr}\left\{\gamma_\mu S_m(K)\gamma_\nu S_m(R)\right\}\nn\\
&= \sum_{f}\frac{ig^2}{2} \int\frac{d^4K}{(2\pi)^4}~ e^{-\frac{k_\perp^2+r_\perp^2}{q_fB}}\sum_{l=0}^\infty\sum_{l'=0}^\infty (-1)^{l+l'}\nn\\
&~~\left(\frac{1}{K_\shortparallel^2-m_f^2-2lq_fB}\right)\left(\frac{1}{R_\shortparallel^2-m_f^2-2l'q_fB}\right)\nn\\
& \times \textsf{Tr}\Bigl\{\gamma_\mu~ D_l(q_fB,K)~\gamma_\nu~ D_{l'}(q_fB,R)\Bigr\},
\label{arbit_self_initial_expression}
\end{align}
where we have used the form of the fermionic propagator from Eq.~\eqref{decomposed_propagator}, replacing the HQ mass $M$ by flavor dependent light quark mass $m_f$. We can further simplify the expression by the virtue of the Hard thermal loop approximation, neglecting the external momenta and the quark mass $m_f$ in the numerator, i.e.
\begin{align}
\Pi_{\mu\nu}^m(Q) &= \sum_{f}\frac{ig^2}{2} \int\frac{d^4K}{(2\pi)^4}~ e^{-\frac{2k_\perp^2}{q_fB}}\sum_{l=0}^\infty\sum_{l'=0}^\infty (-1)^{l+l'}\nn\\
&\left(\frac{1}{K_\shortparallel^2-m_f^2-2lq_fB}\right)\left(\frac{1}{R_\shortparallel^2-m_f^2-2l'q_fB}\right)\nn\\
& \times \textsf{Tr}\Bigl\{\gamma_\mu~ D_l(q_fB,K)~\gamma_\nu~ D_{l'}(q_fB,R)\Bigr\}.
\label{arbit_self_middle_expression}
\end{align}

The nontrivial trace part can be computed by progressing term by term (see Eq.~\eqref{d_l}),
\begin{align}
& \textsf{Tr}\Bigl\{\gamma_\mu ~D_l(K)~\gamma_\nu ~ D_{l'}(R)\Bigr\} = T_{\mn}=\sum_{i=1}^9 (T_i)_{\mn}.
\label{trace_total}
\end{align}
We list the explicit expressions for each of the terms in Appendix \ref{appA}. Evaluation of the trace finally yields the fermionic part of the one-loop gluon self energy to be 
\begin{align}
&\Pi_{\mu\nu}^m(Q) = \sum_{f}\frac{ig^2}{2} \int\frac{d^4K}{(2\pi)^4}~ e^{-\frac{2k_\perp^2}{q_fB}}\sum_{l=0}^\infty\sum_{l'=0}^\infty (-1)^{l+l'}\nn\\
&\frac{\sum_{i=1}^9 (T_i)_{\mn}}{\left(K_\shortparallel^2-m_f^2-2lq_fB\right)\left(R_\shortparallel^2-m_f^2-2l'q_fB\right)}.
\label{arbit_self_final_expression}
\end{align}
Eq.~\eqref{arbit_self_final_expression} along with Eq.~\eqref{Pi_mn_g} give the final expression for the one-loop gluon self energy in a hot and arbitrarily magnetized medium.

After having the expression for the one-loop gluon self energy $\Pi_{\mu\nu}(Q)$,  we can proceed to evaluate the form factors in Eqs.~\eqref{ff_di} one by one. To do that, we need to contract various tensor basis $\Delta_i^{\mn}$ (explicitly given in Eqs.~\eqref{Deltamunu}) with $\Pi_{\mu\nu}(Q)$. We provide the expressions for the four form factors in the next four subsections. For the corresponding thorough steps of contractions, readers should look into Appendix \ref{appB}. 

\subsection{Form factor $d_1$}
We first focus on $d_1$, which can be written as (See Eq.~\eqref{ff_d1})
\begin{align}
d_1 &= \Delta_1^{\mn}(\Pi_{\mn}^{g}+\Pi_{\mn}^m)=d_1^{YM}+d_1^m
\end{align}
where $\Delta_1^{\mn}$ is given in Eq.~\eqref{D1munu}. $d_1^{YM}$ is the contribution for the gluon part, i.e. 
\begin{align}
d_1^{YM}&= \Delta_1^{\mn}\Pi_{\mn}^{g} = \frac{N_cg^2T^2}{3\bar u^2}\left[1-\mathcal{T}_Q(q_0,q)\right].
\end{align}
On the other hand $d_1^m$ is the contribution from the quark loop (See Appendix \ref{appB}), given as
\begin{align}
&d_1^m= \Delta_1^{\mn}\Pi_{\mn}^m = -\sum_f~\frac{g^2 |q_fB|}{2\pi\bar{u}^2}\sum_{l,l'=0}^\infty (-1)^{l+l'}\nn\\
&\left[(\delta_{l,l'}+\delta_{l-1,l'-1})\int\frac{dk_3}{2\pi}\left\{(2-\bar{u}_\sp^2)\Phi_1+(2k_3^2+\right.\right.\nn\\
&\left. (2-\bar{u}_\sp^2)2l|q_fB|)\Phi_2\right\}+\bar{u}_\perp^2(\delta_{l,l'-1}+\delta_{l-1,l'})\int\frac{dk_3}{2\pi}\nn\\
&\left.\left\{\Phi_1+2l|q_fB|\Phi_2\right\}+4\bar{u}^2|q_fB|l\delta_{l,l'}\int\frac{dk_3}{2\pi}\Phi_2\right],\label{coeff_d1m}
\end{align}
where   $\bar u_\sp^2=1-\frac{q_0^2}{Q^2}\left(2-\frac{Q_\sp^2}{Q^2}\right)$, $\bar u_\perp^2 = \frac{q_0^2q_\perp^2}{Q^4}$ and $\bar u^2=-q^2/Q^2$.
In expressing $d_1^m$ we have also used the shorthand notations $\Phi_i$'s for the following frequency sums
\begin{align}
    \Phi_1 &= T\sum_{k_0} \frac{1}{K_\sp^2-m_f^2-2l|q_fB|}\nn\\
    &= T\sum_{k_0} \frac{1}{k_0^2-E_{k_3}^2} =\frac{n_F(E_{k_3})}{E_{k_3}}.
\end{align}
and 
\begin{align}
    \Phi_2 &= T\sum_{k_0} \frac{1}{(K_\sp^2-m_f^2-2l|q_fB|)(R_\sp^2-m_f^2-2l'|q_fB|)}\nn\\
    &= T\sum_{k_0} \frac{1}{(k_0^2-E_{k_3}^2)(r_0^2-E_{r_3}^2)}\nn\\
    &= -\sum_{s_1,s_2=\pm 1}\frac{s_1s_2}{4E_{k_3}E_{r_3}}\left(\frac{1-n_F(s_1E_{k_3})-n_F(s_2E_{r_3})}{q_0-s_1E_{k_3}-s_2E_{r_3}}\right), 
    \label{phi2_initial}
\end{align}
where $E_{k_3} = \sqrt{k_3^2+m_f^2+2l|q_fB|}$ and $E_{r_3} = \sqrt{r_3^2+m_f^2+2l'|q_fB|}$.

At this point, we should make some comment about the LLL approximation. One can see in a straightforward way that from Eq.~\eqref{coeff_d1m}, it is easy to obtain the LLL limit by putting $l=l'=0$, i.e. 
\begin{align}
d_1^{LLL}= d_1^{YM}-\sum_f~\frac{g^2 |q_fB|}{2\pi\bar{u}^2}\int\frac{dk_3}{2\pi}\left[\Phi_1+2k_3^2\Phi_2\right].\label{coeff_d1_lll}
\end{align}
This is similar to what has been obtained in Ref.~\cite{Fukushima:2015wck}, where the authors further exploited HTL approximations within $\Phi_2$ to simplify the expression for $d_1$ and express it in terms of the medium modified Debye mass $m_D$.

\subsection{Form factor $d_2$}

Focusing on $d_2$, we have (See Eqs.~\eqref{ff_d2} and \eqref{D2munu})
\begin{align}
d_2 &= \Delta_2^{\mn}(\Pi_{\mn}^{g}+\Pi_{\mn}^m)=d_2^{YM}+d_2^m.
\end{align}
The pure-glue part $d_2^{YM}$ yields
\begin{align}
d_2^{YM} &= \Delta_2^{\mn}\Pi_{\mn}^{g} =\frac{N_cg^2T^2}{3}\frac{1}{2}\left[\frac{q_0^2}{q^2}-\frac{Q^2}{q^2}\mathcal{T}_Q(q_0,q)\right].
 \label{coeff_d2YM}
\end{align}
For the quark part, we obtain the final expression as 
\begin{align}
&d_2^m= \Delta_2^{\mn}\Pi_{\mn}^m = \sum_f~\frac{g^2 |q_fB|}{2\pi}\sum_{l,l'=0}^\infty (-1)^{l+l'}\nn\\
&\left[(\delta_{l,l'-1}+\delta_{l-1,l'})\int\frac{dk_3}{2\pi}\left\{\Phi_1+2l|q_fB|\Phi_2\right\}\right.\nn\\
&\left.+4|q_fB|l\delta_{l,l'}\int\frac{dk_3}{2\pi}\Phi_2\right].
\label{coeff_d2m}
\end{align}

\subsection{Form factor $d_3$}

Similar to $d_1$ and $d_2$, we can break down $d_3$ into two parts (See Eqs.~\eqref{ff_d3} and \eqref{D3munu}), i.e.
\begin{align}
d_3 &= \Delta_3^{\mn}(\Pi_{\mn}^{g}+\Pi_{\mn}^m)=d_3^{YM}+d_3^m
\end{align}
where the pure glue part yields
\begin{align}
d_3^{YM}&=\Delta_3^{\mn}\Pi_{\mn}^{g} = \frac{N_cg^2T^2}{3}\frac{1}{2}\left[\frac{q_0^2}{q^2}-\frac{Q^2}{q^2}\mathcal{T}_Q(q_0,q)\right].
\end{align}
For the quark part we can subsequently write down 
\begin{align}
&d_3^m= \Delta_3^{\mn}\Pi_{\mn}^m = -\sum_f~\frac{g^2 |q_fB|}{2\pi\bar{n}^2}\sum_{l,l'=0}^\infty (-1)^{l+l'}\nn\\
&\left[-\bar{n}_\sp^2(\delta_{l,l'}+\delta_{l-1,l'-1})\int\frac{dk_3}{2\pi}\left\{\Phi_1+(2k_3^2\right.\right.\nn\\
&\left.+2l|q_fB|)\Phi_2\right\}+\bar{n}_\perp^2(\delta_{l,l'-1}+\delta_{l-1,l'})\int\frac{dk_3}{2\pi}\nn\\
&\left.\left\{\Phi_1+2l|q_fB|\Phi_2\right\}+4\bar{n}^2|q_fB|l\delta_{l,l'}\int\frac{dk_3}{2\pi}\Phi_2\right],
\label{coeff_d3m}
\end{align}
where  $\bar n_\sp^2 = -\frac{q_\perp^4}{q^4}$, $\bar n_\perp^2 = \frac{q_3^2q_\perp^2}{Q^4}$ and $\bar n^2=-\frac{q_{\perp}^2}{q^2}$.

\subsection{Form Factor $d_4$}

Finally for the last form factor $d_4$ (See Eqs.~\eqref{ff_d4} and \eqref{D4munu}) the Yang-Mills contribution vanishes as 
\begin{align}
d_4^{YM} &= \Delta_4^{\mn}\Pi_{\mn}^{g} = 0,
\end{align}
and we are only left with the quark loop contribution, i.e. $d_4=d_4^m$. Subsequently $d_4^m$ comes out to be
\begin{align}
&d_4^m= \frac{1}{2}\Delta_4^{\mn}\Pi_{\mn}^{m} = -\sum_f~\frac{g^2 |q_fB|}{2\pi\sqrt{\bar{u}^2}\sqrt{\bar{n}^2}}\sum_{l,l'=0}^\infty (-1)^{l+l'}\nn\\
&\left[(\delta_{l,l'}+\delta_{l-1,l'-1})\int\frac{dk_3}{2\pi}\left\{2k_3\bar{n}^2\Phi_3-(\bar{n}\cdot\bar{u})_\sp \times \right. \right.\nn\\ 
& \left.(\Phi_1+2l|q_fB|\Phi_2) \right\}+(\bar{n}\cdot\bar{u})_\perp(\delta_{l,l'-1}+\delta_{l-1,l'})\int\frac{dk_3}{2\pi}\nn\\
&\left.\left\{\Phi_1+2l|q_fB|\Phi_2\right\}\right],
\label{coeff_d4m}
\end{align}
where $(\bar n \cdot \bar u)_\sp = (\bar n \cdot \bar u)_\perp = \frac{q_0q_3q_\perp^2}{q^2Q^2}$.

Here we have introduced a third frequency sum $\Phi_3$ as 
\begin{align}
    \Phi_3 &= T\sum_{k_0} \frac{k_0}{(K_\sp^2-m_f^2-2l|q_fB|)(R_\sp^2-m_f^2-2l'|q_fB|)}\nn\\
    &= T\sum_{k_0} \frac{k_0}{(k_0^2-E_{k_3}^2)(r_0^2-E_{r_3}^2)}\nn\\
    &= -\sum_{s_1,s_2=\pm 1}\frac{s_2}{4E_{r_3}}\left(\frac{1-n_F(s_1E_{k_3})-n_F(s_2E_{r_3})}{q_0-s_1E_{k_3}-s_2E_{r_3}}\right).
\end{align}

Now that we have all the expressions for the individual form factors $d_i$'s, the construction of the effective HTL gluon propagator $\mathcal{D}^{\mn}(Q)$ involving quarks residing on any arbitrary Landau levels seems complete (See Eq.~\eqref{gauge_prop}). In the next section we will proceed with the next steps, i.e. we will compute the HQ scattering rate $\Gamma(P)$ in a hot magnetized medium, utilizing the HQ effective self energy $\Sigma(P)$.


\section{HQ scattering rate ($\Gamma$) in a magnetized medium}
\label{sec4}
We start this section with the expression for the HQ effective self energy. Using Eq.~\eqref{decomposed_propagator} and Eq.~\eqref{gauge_prop} in Eq.~\eqref{htl_eff_self_energy}, we can write down the HQ effective self energy as 
\begin{align}
    \Sigma(P) &= ig^2\sum_{l=0}^{\infty}(-1)^l\int\frac{d^4Q}{(2\pi)^4} \frac{e^{-{k_\perp^2}/{|q_fB|}}}{K_\sp^2-M^2-2l|q_fB|} \nn\\
&\times\sum\limits_{i=1}^4\mathcal{J}_i\Delta_i^{\mn}\gamma_\mu D_l(q_fB,K)\gamma_\nu,
\label{HQ_effective_self_energy2}
\end{align}
which also makes it quite evident that we have chosen a gauge with vanishing gauge parameter, i.e. $\xi=0$ in Eq.~\eqref{gauge_prop}.

Now to evaluate the scattering rate using Eq.~\eqref{interaction_rate2}, one needs to evaluate the imaginary part of $\Sigma(P)$ and perform a trace, i.e.  $\Tr\left[(\slashed{P}+M) {\rm Im}~\Sigma(P)\right]$. These two actions can commute with each other. Hence we will first perform the trace and then evaluate the imaginary part to obtain the final expression for the HQ scattering rate in a magnetized medium. From Eq.~\eqref{HQ_effective_self_energy2}, one can write the trace as :
\begin{align}
&\Tr\left[(\slashed{P}+M)~\Sigma(P)\right] = ig^2\sum_{l=0}^{\infty}(-1)^l\nn\\
&~~\int\frac{d^4Q}{(2\pi)^4} \frac{e^{-{k_\perp^2}/{|q_fB|}}}{K_\sp^2-M^2-2l|q_fB|} \nn\\
&\times\sum\limits_{i=1}^4\mathcal{J}_i~\Tr\left[(\slashed{P}+M)\Delta_i^{\mn}\gamma_\mu D_l(q_fB,K)\gamma_\nu\right].\label{trace_HQ_self_energy}
\end{align}
We can now proceed to evaluate the individual traces $\Tr\left[(\slashed{P}+M)\Delta_i^{\mn}\gamma_\mu D_l(q_fB,K)\gamma_\nu\right]$. To this end, we will break down each of the traces into two parts, i.e. $q_0$ independent $A_i$'s and $q_0$ dependent $B_i$'s, i.e.
\begin{align}
    \Tr\left[(\slashed{P}+M)\Delta_i^{\mn}\gamma_\mu D_l(q_fB,K)\gamma_\nu\right] = A_i + B_i(q_0).
    \label{traces_Ai}
\end{align}
The reason for this arrangement can be understood as follows :
\begin{enumerate}
    \item We are eventually interested in the imaginary part of $\Sigma(P)$. It has been explicitly shown in Ref.~\cite{Braaten:1991jj} for $eB=0$ and in Ref.~\cite{Bandyopadhyay:2021zlm} for $eB\neq 0$ that $q_0$ dependent terms will not contribute in the imaginary part of $\Sigma(P)$. 
    \item We will work on the small energy transfer limit of HQ scattering, which essentially requires $q_0\rightarrow 0$. Within this limit, various form factors $d_i$s undergoes further simplifications which we eventually use in our numerical evaluation. These simplifications have been explicitly discussed in Appendix~\ref{appC}.    
\end{enumerate}
Hence we can safely neglect the $q_0$ dependent pieces $B_i$'s to proceed further into the calculation. Explicit evaluations of the traces along with the expressions for $A_i$'s are given in Appendix~\ref{appD}.

To compute the sum over $q_0$, we introduce the spectral representations for the HQ and effective gluon propagators respectively by using the following relations  
\begin{align}
&\frac{1}{K_\sp^2-M^2-2l|q_fB|} =-\frac{1}{2E'_\sp}\times\nn\\
& \int\limits_0^{1/T}d\tau' e^{k_0\tau'} \left[(1-n_F(E'_\sp))e^{-E'_\sp\tau'} - n_F(E'_\sp)e^{E'_\sp\tau'} \right],\label{HQ_prop_spectral}\\
&\mathcal{J}_i = - \int\limits_0^{1/T}d\tau~ e^{q_0\tau} \int\limits_{-\infty}^{+\infty}~ d\omega~ \rho_i(\omega,q)~\left[1+n_B(\omega)\right]~e^{-\omega\tau}.\label{eff_gluon_prop_spectral}
\end{align}
For the spectral representation of the HQ propagator, we have used the dispersion relation $E'_\sp = \sqrt{k_3^2+M^2+2l|q_fB|}$. On the other hand for the spectral representation of the effective gluon propagator we have defined the spectral functions as
\begin{align}
&\rho_i(\omega,q) = \frac{1}{\pi} {\rm Im}\left(\mathcal{J}_i\Big |_{q_0 = \omega +i\epsilon}\right),\label{spec_func_gen}
\end{align}
which extract the imaginary parts of the corresponding coefficients $\mathcal{J}_i$. Explicit expressions of these spectral functions are given in Appendix \ref{appE}.

Once we use Eqs.~\eqref{traces_Ai}, \eqref{HQ_prop_spectral} and \eqref{eff_gluon_prop_spectral} in Eq.~\eqref{trace_HQ_self_energy}, the sum over $q_0$ can be evaluated from the combination of the integrals over $\tau$ and $\tau'$, using\footnote{Similar formulae have been used in our earlier work~\cite{Bandyopadhyay:2021zlm}, which contain typos of missing a factor of $T$.} 
\begin{align}
T\sum_{q_0} e^{q_0(\tau-\tau')} =& \delta (\tau-\tau').
\end{align}
This subsequently yields
\begin{align}
&\Tr\left[(\slashed{P}+M)~\Sigma(P)\right] =~ ig^2\sum_{l=0}^{\infty}(-1)^l\nn\\
&\int\frac{d^4Q}{(2\pi)^4}\frac{e^{-{k_\perp^2}/{|q_fB|}}}{K_\sp^2-M^2-2n|q_fB|}  \sum\limits_{i=1}^4 \mathcal{J}_i~A_i \nn\\
&= -g^2\sum_{n=0}^{\infty}(-1)^l\sum\limits_{i=1}^4\int\frac{d^3q}{(2\pi)^3} e^{-{k_\perp^2}/{|q_fB|}}\nn\\
&\int\limits_{-\infty}^{+\infty} d\omega\left[1+n_B(\omega)\right]\frac{\rho_i(\omega,q)}{2E'_\sp} A_i~P_1,
\end{align}
where 
\begin{align}
P_1 =& \int\limits_0^{1/T}d\tau'\int\limits_0^{1/T}d\tau ~e^{p_0\tau'-\omega\tau} \delta (\tau-\tau')\nn\\
&~\times\left[(1-n_F(E'_\sp))e^{-E'_\sp\tau'} - n_F(E'_\sp)e^{E'_\sp\tau'}\right]\nn\\
=&-\sum_{\sigma=\pm 1} \frac{\sigma~n_F(\sigma E'_\sp)}{p_0+\sigma E'_\sp-\omega}\left(e^{(p_0+\sigma E'_\sp-\omega)/T}-1\right).
\label{P1_final}
\end{align}

After this we can proceed to compute the discontinuity by writing down the evaluation for the trace, now incorporating the imaginary part, as  
\begin{align}
&\Tr\left[(\slashed{P}+M)~{\rm Im}\Sigma(p_0+i\epsilon, \vec{p})\right] \nn \\
=&~ \pi g^2\sum_{l=0}^{\infty}(-1)^l\left(e^{-E/T}+1\right)\sum\limits_{i=1}^4 \int\frac{d^3q}{(2\pi)^3} e^{-{k_\perp^2}/{|q_fB|}}\nn\\
&\times \int\limits_{-\infty}^{+\infty} d\omega\left[1+n_B(\omega)\right]\frac{\rho_i(\omega,q)A_i}{2E'_\sp}\nn\\
&\times\sum_{\sigma=\pm 1} \sigma~n_F(\sigma E'_\sp)~\delta(E + \sigma E'_\sp - \omega).
\label{trace_imag_final}
\end{align}

Finally using Eq.~\eqref{trace_imag_final} in Eq.~(\ref{interaction_rate2}), we can obtain the scattering rate $\Gamma(P)$ (for a particular HQ flavor) as 
\begin{align}
\Gamma(P) =& -\frac{\pi g^2}{2E} \sum_{l=0}^{\infty}(-1)^l\sum\limits_{i=1}^4 \int\frac{d^3q}{(2\pi)^3} e^{-{k_\perp^2}/{|q_fB|}}\nn\\
&\times \int\limits_{-\infty}^{+\infty} d\omega\left[1+n_B(\omega)\right]\frac{\rho_i(\omega,q)A_i}{2E'_\sp}\nn\\
&\times\sum_{\sigma=\pm 1} \sigma~n_F(\sigma E'_\sp)~\delta(E + \sigma E'_\sp - \omega).
\end{align}

We can further simplify the expression for the scattering rate a bit further using the scale hierarchy $M\gg \sqrt{eB}, T$. As $E \sim E'_\sp \sim M$, so the delta function $\delta(E + E'_\sp-\omega)$ cannot contribute for $\omega \leq T$. Also, the Fermi-Dirac distribution $n_F(E'_\sp)$ will be exponentially suppressed. These changes subsequently simplify the expression of the scattering rate as 
\begin{align}
&\Gamma(P) = \frac{\pi g^2}{2E} \sum_{l=0}^{\infty}(-1)^l\sum\limits_{i=1}^4 \int\frac{d^3q}{(2\pi)^3} e^{-{k_\perp^2}/{|q_fB|}}\nn\\
&\int\limits_{-\infty}^{+\infty} d\omega\left[1+n_B(\omega)\right]\frac{\rho_i(\omega,q)A_i}{2E'_\sp} \delta(E - E'_\sp - \omega).
\label{scattering_rate_final}
\end{align}
Eq.~\eqref{scattering_rate_final} is the final expression for the HQ scattering rate in a hot and arbitrarily magnetized medium which will be used in the next section to write down the HQ momentum diffusion coefficients.


\section{HQ momentum diffusion coefficients in a magnetized medium}
\label{sec5}

In this section we write down the final expressions for the HQ momentum diffusion coefficients $\kappa_i$'s within the most general scenario of a hot magnetized medium from Eqs.~\eqref{coeffs_case1} and \eqref{coeffs_case2} using the already computed scattering rate in the previous section. Beyond the static limit of the HQ we have considered two simple cases, HQ moving parallel (case 1) or perpendicular (case 2) to the direction of the magnetic field. Following we provide explicit expressions for various HQ momentum diffusion coefficients within these two cases respectively.

\subsection{case 1 : $\bf{v} \sp \bf{B}$}
\label{case1_exprs}

As the magnetic field is considered to be along the $z$ direction, for this case we only have a nonzero $p_3$ whereas $p_1=p_2=0$, resulting $E=\sqrt{p_3^2+M^2}$. The scattering rate is same as Eq.~\eqref{scattering_rate_final} with $A_i$'s replaced by $A_i^{(1)} = A_i(p_1=p_2=0)$. Explicit expressions for $A_i^{(1)}$'s are given in Appendix~\ref{appD}. Subsequently the transverse momentum diffusion coefficient for this case will be given using Eq.~\eqref{coeffs_case1} as 
\begin{align}
&\kappa_T(p_3) = \frac{\pi g^2T}{8E} \sum_{l=0}^{\infty}(-1)^l \sum\limits_{i=1}^4 \int\frac{d^3q}{(2\pi)^3} q_\perp^2 e^{-{q_\perp^2}/{|q_fB|}}\nn\\
&\int\limits_{-\infty}^{+\infty} d\omega \frac{\rho_i(\omega,q)A_i^{(1)}}{\omega E'_\sp} \delta(E - E'_\sp - \omega),
\label{kappaT_case1_final}
\end{align}
where we replace the factor $(1+n_B(\omega))$ with $\frac{T}{\omega}$, owing to the small energy transfer limit.

 Similarly the longitudinal momentum diffusion coefficient will be given as 
\begin{align}
&\kappa_L(p_3) = \frac{\pi g^2T}{4E} \sum_{l=0}^{\infty}(-1)^l \sum\limits_{i=1}^4 \int\frac{d^3q}{(2\pi)^3} q_3^2 ~e^{-{q_\perp^2}/{|q_fB|}}\nn\\
&\int\limits_{-\infty}^{+\infty} d\omega\frac{\rho_i(\omega,q)A_i^{(1)}}{\omega E'_\sp} \delta(E - E'_\sp - \omega).
\label{kappaL_case1_final}
\end{align}

\subsection{case 2 : $\bf{v} \perp \bf{B}$}

For this case we have nonzero $p_1$ and/or $p_2$ whereas $p_3=0$. Hence $E=\sqrt{p_\perp^2 + M^2}$ and $E'_\sp = \sqrt{q_3^2 + M^2 + 2l|q_fB|}$. The scattering rate expression in Eq.~\eqref{scattering_rate_final} will again have modifications in $A_i$'s which will now be replaced by $A_i^{(2)} = A_i (p_3=0)$. Hence using Eq.~\eqref{scattering_rate_final} and Eq.~\eqref{coeffs_case2}, we can straightway write down the expressions for the momentum diffusion coefficients as
\begin{align}
\kappa_j(p) =& \frac{\pi g^2T}{4E} \sum_{l=0}^{\infty}(-1)^l \sum\limits_{i=1}^4 \int\frac{d^3q}{(2\pi)^3} q_j^2~ e^{-{k_\perp^2}/{|q_fB|}}\nn\\
&\int\limits_{-\infty}^{+\infty} d\omega\frac{\rho_i(\omega,q)A_i^{(2)}}{\omega E'_\sp} \delta(\omega-E+E'_\sp).
\end{align}
Again we have replaced $(1+n_B(\omega))$ with $\frac{T}{\omega}$ due to the small energy transfer limit. Explicit expressions for $A_i^{(2)}$ are given in Appendix~\ref{appD}.


\section{Results I : Discussions on the static limit}
\label{sec6}

Before going into the estimations of the HQ momentum diffusion coefficients beyond the static limit of the HQ, in this section we will discuss the results considering the static limit of the HQ. From our general expression given in section~\ref{sec5}, we can readily revert back to the static limit considering the HQ momentum $\vec{p}$ (vis-a-vis velocity $\vec{v}$) to be vanishing, i.e. $E\approx M$. Firstly we notice that without the anisotropy generated by $\vec{v}$, $\vec{v}\perp\vec{B}$ case will have no contribution in the static limit. There will be only one case with $\vec{B}$ providing the anisotropic direction, which is essentially the $\vec{v}\sp\vec{B}$ case with $v\rightarrow 0$. The simplified expressions for $A_i$'s within the static limit are given in Appendix~\ref{appD}, from where we can see that the only non vanishing contributions comes from $A_1$ and $A_4$. Out of these two, $A_4$ term eventually does not contribute because of the vanishing of its associated spectral contribution $\rho_4$ (See Appendix~\ref{appE}). The reason for this is explained in Appendix~\ref{appC} which shows $d_4\approx 0$ in the small energy transfer limit. Hence incorporating all these factors we can write down the expressions for $\kappa_T$ and $\kappa_L$ within the static limit from Eqs.~\eqref{kappaT_case1_final} and \eqref{kappaL_case1_final}, as
\begin{align}
\kappa_T^{(s)} &= \frac{\pi g^2T}{8M} \sum_{l=0}^{\infty}\frac{(-1)^l}{\sqrt{M^2+2l|q_fB|}} \int\frac{d^3q}{(2\pi)^3} q_\perp^2 e^{-{q_\perp^2}/{|q_fB|}}\nn\\
&~\times \left[\frac{1}{\omega}\rho_1^{(s)}(\omega,q)~A_1^{(s)} \right]_{\omega\rightarrow 0},
\label{kappaT_static}
\end{align}

\begin{align}
\kappa_L^{(s)} &= \frac{\pi g^2T}{4M} \sum_{l=0}^{\infty}\frac{(-1)^l}{\sqrt{M^2+2l|q_fB|}} \int\frac{d^3q}{(2\pi)^3} q_3^2 ~e^{-{q_\perp^2}/{|q_fB|}}\nn\\
&~\times \left[\frac{1}{\omega} \rho_1^{(s)}(\omega,q)~A_1^{(s)} \right]_{\omega\rightarrow 0}.
\label{kappaL_static}
\end{align}

\begin{figure*}[t]
\begin{center}
\includegraphics[scale=0.5]{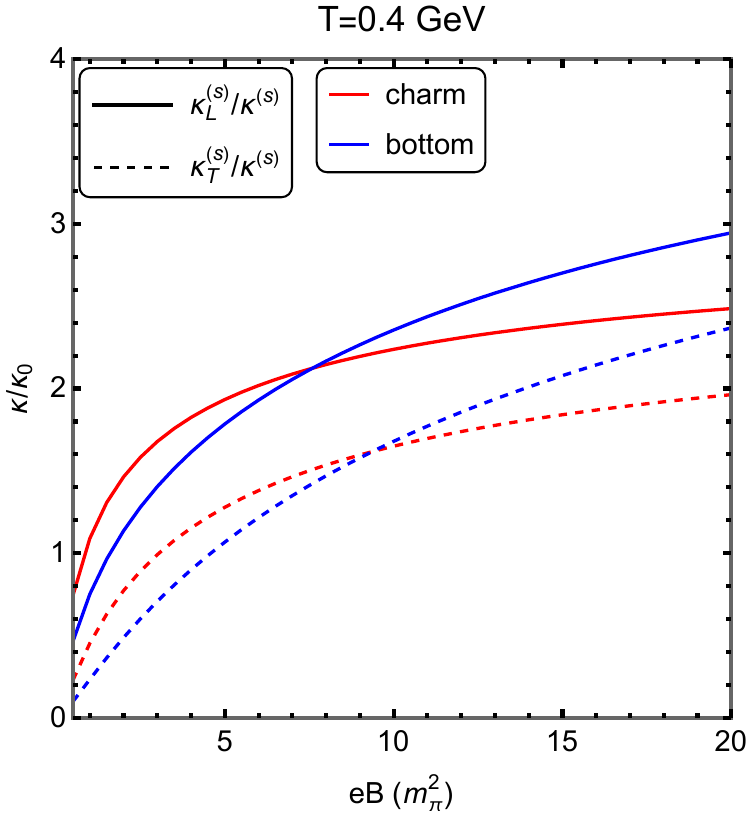}
\hspace{1cm}
\includegraphics[scale=0.5]{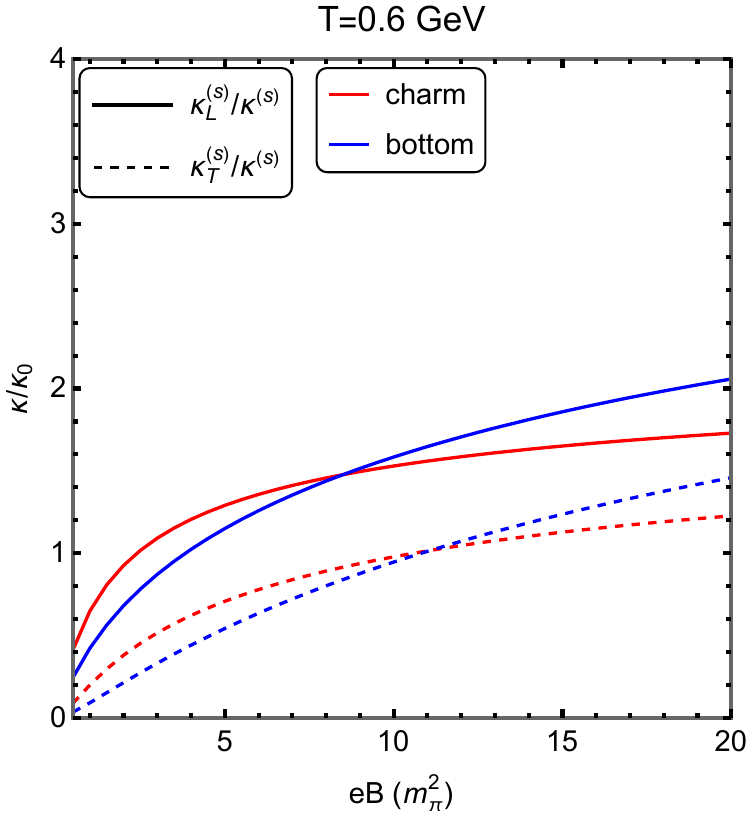}
\caption{The magnetized medium modified exact results ($\kappa$) has been scaled with respect to the $eB=0$ result ($\kappa_0$), variation of which with respect to $eB$ has been shown for longitudinal (solid lines) and transverse (dashed lines) HQ momentum diffusion coefficients within the static limit of both charm (red curves) and bottom (blue curves) quarks. Comparison has been done for two different values of temperatures i.e. $T=0.4$ GeV (left panel) and $T=0.6$ GeV (right panel). Charm and bottom quark masses $M$ are specified in the text.} 
\label{kappa_ratio_static_wzB}
\end{center}
\end{figure*}

Here $A_1^{(s)}$ and $\rho_1^{(s)}$ are the simplified expressions for $A_1$ and $\rho_1$ within the static and small energy transfer limit of the HQ, given in Appendices \ref{appD} and \ref{appE} respectively (See Eqs.~\eqref{A1_static} and \eqref{rho1_static}). Putting those in Eqs.~\eqref{kappaT_static} and \eqref{kappaL_static} we obtain 
\begin{align}
&\kappa_T^{(s)} = \sum_{l=0}^{\infty}\frac{(-1)^{l}\pi g^2TM }{\sqrt{M^2+2l|q_fB|}} \int\frac{d^3q}{(2\pi)^3} q_\perp^2 e^{-{q_\perp^2}/{|q_fB|}}\nn\\
&\left[\frac{\left(\frac{1}{q}(m_D^g)^2+\delta(q_3)\sum_f\delta m_{D,f}^2\right)(L_l(\xi_q^\perp) - L_{l-1}(\xi_q^\perp))}{2(q^2+(m_D')^2)^2}\right],
\label{kappaT_static_final}\\
&\kappa_L^{(s)} = \sum_{l=0}^{\infty}\frac{(-1)^{l}~2\pi g^2TM }{\sqrt{M^2+2l|q_fB|}} \int\frac{d^3q}{(2\pi)^3} q_3^2 e^{-{q_\perp^2}/{|q_fB|}}\nn\\
&\left[\frac{(m_D^g)^2(L_l(\xi_q^\perp) - L_{l-1}(\xi_q^\perp))}{2q(q^2+(m_D')^2)^2}\right],
\label{kappaL_static_final}
\end{align}
where $m_D'$ is the full magnetized medium modified QCD Debye mass which consists of the pure glue part $m_D^g$ and the magnetic field modified correction $\delta m_D$, given as
\begin{align}
(m_D')^2 &= (m_D^g)^2 + \sum_f \delta m_{D,f}^2, \\
    (m_D^g)^2 &= \frac{N_c g^2T^2}{3},\\
    \delta m_{D,f}^2 &= -\frac{g^2 |q_fB|}{4\pi^2}\sum_{l=0}^\infty (2-\delta_{l,0})\int dk_3~ \frac{\partial n_F(E_{k_3})}{\partial E_{k_3}},\nn\\
    &= \frac{g^2 |q_fB|}{4\pi^2T}\sum_{l=0}^\infty (2-\delta_{l,0})\int dk_3~ n_F(1-n_F).
\end{align}

From Eqs.~\eqref{kappaT_static_final} and \eqref{kappaL_static_final}, one can also easily revert back to the lowest Landau level limit by putting $l=0$ and see that it matches with previous results within the same limit~\cite{Fukushima:2015wck}. 

\begin{figure*}[t]
\begin{center}
\includegraphics[scale=0.5]{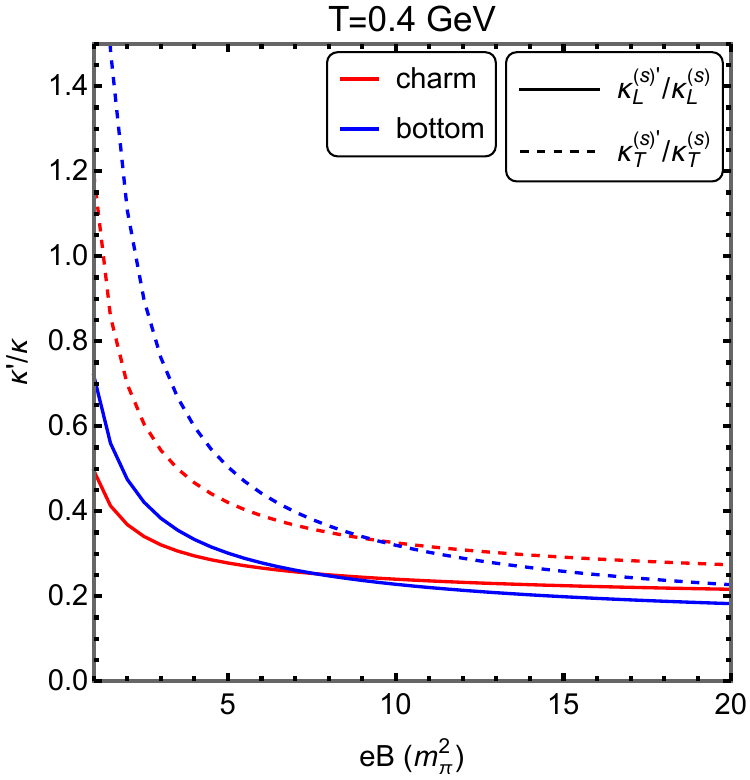}
\hspace{1cm}
\includegraphics[scale=0.5]{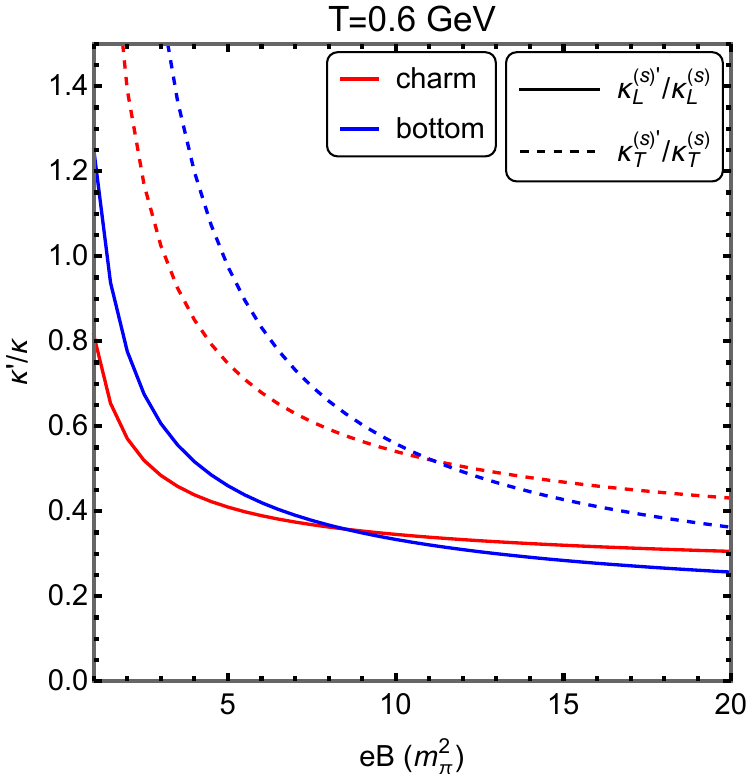}
\caption{Variation of the ratio between the Debye mass approximated results ($\kappa'$) and the exact results ($\kappa$) with respect to $eB$ has been shown for longitudinal (solid lines) and transverse (dashed lines) HQ momentum diffusion coefficients within the static limit of both charm (red curves) and bottom (blue curves) quarks. Comparison has been done for two different values of temperatures i.e. $T=0.4$ GeV (left panel) and $T=0.6$ GeV (right panel). Charm and bottom quark masses $M$ are specified in the text.} 
\label{kappa_ratio_static}
\end{center}
\end{figure*}

Before moving further, let us briefly mention the static limit result for vanishing magnetic field. In this case, since there is no spatial anisotropy in the system, there will be a single momentum diffusion coefficient $\kappa^{(s)}$. To provide an expression for $\kappa^{(s)}$, we can start with the $eB=0$ expression for the HQ scattering rate explored in Refs~\cite{Braaten:1991we,Beraudo:2009pe,Bandyopadhyay:2021zlm}, i.e. 
\begin{align}
\Gamma_{eB=0}&= 2\pi g^2 \!\!\int\!\!\!\frac{d^3q}{(2\pi)^3}\!\!\int\limits_{-\infty}^{+\infty}\!\!d\omega[1+n_B(\omega)]\delta(\omega -\vec{v}\cdot\vec{q})\nn\\
&\left[ \rho_L(\omega,q)+\rho_T(\omega,q)(v^2-(\vec{v}\cdot\hat{q})^2)\right].
\label{sr_zeroB}
\end{align}
Considering the static limit of the HQ and small energy transfer, we further simplify the HQ scattering rate as 
\begin{align}
\Gamma_{eB=0}^{(s)}&= 2\pi g^2T \!\!\int\!\!\!\frac{d^3q}{(2\pi)^3}\!\!\left[\frac{1}{\omega} \rho_L(\omega,q)\right]_{\omega\rightarrow 0},\nn\\
&= 2\pi g^2T \!\!\int\!\!\!\frac{d^3q}{(2\pi)^3}\!\!\left[\frac{\frac{1}{2q}m_D^2}{(q^2+m_D^2)^2}\right],
\label{sr_zeroB_static}
\end{align}
where we have used the expression of the HTL longitudinal spectral function $\rho_L$~\cite{Braaten:1991we,Beraudo:2009pe,Mustafa:2022got,Bellac:2011kqa}, according to our definition of spectral function (i.e. Eq.~\eqref{spec_func_gen}). Hence the momentum diffusion coefficient $\kappa$ for $eB=0$ and within the static limit can be straightaway expressed as
\begin{align}
    \kappa^{(s)} &=  2\pi g^2T \!\!\int\!\!\!\frac{d^3q}{(2\pi)^3}\!\!\left[\frac{q~m_D^2}{2(q^2+m_D^2)^2}\right].
\end{align}

In Fig.~\ref{kappa_ratio_static_wzB} we show our most general static limit results for magnetized medium, by varying the ratio $\kappa_{L/T}^{(s)}/\kappa^{(s)}$ with respect to the external magnetic field, with $\kappa^{(s)}$ being the zero magnetic field value of the single momentum diffusion coefficient. Evaluation of $\kappa^{(s)}$ has been done using a temperature dependent UV cutoff $q_{max}$, which will be discussed later. We chose two relatively higher temperatures, i.e. $T=0.4$ and $T=0.6$ GeV, to respect the HTL approximation applied throughout the calculation. It can be noticed from the plots that for lower values of $eB$, the rate of increase for the values of both longitudinal (solid curves) and transverse (dashed curves) momentum diffusion coefficients have been relatively larger than that for higher values of $eB$. This trend is more visible for charm quarks (red curves), which also results in a crossover between the charm and bottom quark (blue) curves. The values of $\kappa_L$ dominate the values of $\kappa_T$ for both static charm and bottom quarks, throughout the range of magnetic field presented in the plots of Fig.~\ref{kappa_ratio_static_wzB}. Finally, with the increase in temperature, the overall value of the ratio decreases, as expected because of the competing scales $eB$ and $T$. 

Next we discuss an alternate procedure to incorporate the effect of the magnetic field in the HQ scattering rate and HQ momentum diffusion coefficients. In this method, all the medium effects are channeled through the medium modified Debye screening mass. Since we have already discussed the static limit scattering rate for $eB=0$, to incorporate the magnetic field effect within the HQ scattering rate, we can straightaway plug the magnetic field modified Debye mass $m_D'(eB,T)$ in Eq.~\eqref{sr_zeroB_static}, and write down the corresponding momentum diffusion coefficients (again considering $\vec{B}=B\hat{z}$), as
\begin{align}
    \kappa_T^{(s)'} &= \pi g^2T \!\!\int\!\!\!\frac{d^3q}{(2\pi)^3}\!\!\left[\frac{q_\perp^2 (m_D')^2}{2q(q^2+(m_D')^2)^2}\right],\label{kappaT_static_alt}\\
    \kappa_L^{(s)'} &= 2\pi g^2T \!\!\int\!\!\!\frac{d^3q}{(2\pi)^3}\!\!\left[\frac{q_3^2 (m_D')^2}{2q(q^2+(m_D')^2)^2}\right]\label{kappaL_static_alt}.
\end{align}

We are finally in a position to both qualitatively and quantitatively compare these two alternate procedure. One can immediately notice the structural similarities between $\kappa$'s and $\kappa'$'s. To make this similarity more prominent to our readers, let us explicitly write down the LLL case for $\kappa_L^{(s)}$, a further simplified expression from Eq.~\eqref{kappaL_static_final}, which reads as
\begin{align}
   \kappa_L^{(s)}\Big|_{LLL} = 2\pi g^2T \int\frac{d^3q}{(2\pi)^3} e^{-{q_\perp^2}/{|q_fB|}}\left[\frac{ q_3^2(m_D^g)^2}{2q(q^2+(m_D')^2)^2}\right].\label{kappaL_static_LLL}
\end{align}
One can immediately notice the striking similarities between the HQ momentum diffusion coefficients presented in Eq.~\eqref{kappaL_static_alt} and Eq.~\eqref{kappaL_static_LLL}.

But looks can be deceiving. When one carefully compares between Eqs.~\eqref{kappaT_static_alt}-\eqref{kappaL_static_alt} and Eqs.~\eqref{kappaT_static_final}-\eqref{kappaL_static_final},\eqref{kappaL_static_LLL}, the significant differences between the two results can be observed, which we list down bellow.
\begin{enumerate}
    \item First and foremost, the structural anisotropy in presence of the magnetic field is not captured in Eqs.~\eqref{kappaT_static_alt}-\eqref{kappaL_static_alt}, which basically means they produce similar values for $\kappa_L$ and $\kappa_T$ at a particular value of the temperature and the magnetic field. But as the exact results (Eqs.~\eqref{kappaT_static_final}-\eqref{kappaL_static_final}) suggest, this is not the case. We explain this issue further in the following point. 

    \item From Eq.~\eqref{kappaL_static_final} one can see that the quark loop contributions for $\kappa_L$ coming from different Landau levels eventually vanish within the static limit because of the factor $\delta(q_3)$, i.e. vanishing longitudinal momentum transfer. This is physically related to the corresponding kinematics constrained by the HTL approximation which neglects the quark mass in leading order. The only non vanishing contribution to $\kappa_L$ comes from the scatterings of the hard thermal gluons. These physical subtleties have not been captured in Eq.~\eqref{kappaL_static_alt}, when one just assumes the modification solely through the Debye mass.
    
    \item Considering the modifications of the HQ propagator in a magnetized medium provide us with the factor $e^{-{q_\perp^2}/{|q_fB|}}$, because of which we do not require any hard UV momentum cutoff for the $q$ integration in the exact procedure. On the other hand, medium modified Debye mass does not provide such soft UV momentum cutoff and one has to put an upper limit $q_{max}$ to get a finite value out of the mediating gluon momenta integration. In Ref~\cite{Beraudo:2009pe}, the authors have estimated this value of $q_{max}$, comparing their results with QCD kinetic calculation. Similar estimations of $q_{max}$ can be tried in magnetized medium through a fitting procedure, comparing the results between $\kappa$ and $\kappa'$. 

    \item There is no explicit HQ mass dependence in the expressions of the Debye mass approximated momentum diffusion coefficients, as opposed to the exact expressions for arbitrary Landau levels. 
\end{enumerate}

In Fig.~\ref{kappa_ratio_static}, we have shown further quantitative comparisons between the results generated from the exact procedure $\kappa$ and the Debye mass approximated procedure $\kappa'$. For evaluating UV finite values of $\kappa'$, we have adopted a similar form of $q_{max}$, as in Ref~\cite{Beraudo:2009pe}, just replacing the temperature dependent one-loop running coupling $g(T)$ by the magnetized medium modified $g(T,eB)$~\cite{Ayala:2014uua,Ayala:2016bbi,Ayala:2018wux,Ayala:2019nna}, i.e. $q_{max}=3.1 T g(T,eB)^{1/3}$. For evaluating the magnetic field induced correction of the Debye mass also we have used $g(T,eB)$, instead of $g(T)$~\cite{Bandyopadhyay:2021zlm}. Fig.~\ref{kappa_ratio_static} shows the variation of the ratio $\kappa'/\kappa$ with respect to magnetic field for two different values of temperature, $T=0.4$ and $0.6$ GeV. For both the longitudinal and transverse components of charm/bottom quark, the basic feature is similar, the Debye mass approximated results underestimate the exact results for larger values of $eB$ and overestimate them for smaller values of $eB$. Both the effects are more prominent in the case of bottom quarks, because of their heavier mass ($M_b = 4.18$ GeV) compare to charm quark ($M_c = 1.27$ GeV). Also one can notice that the values of the ratio of transverse components ($\kappa_T'/\kappa_T$) are larger than that of the longitudinal components ($\kappa_L'/\kappa_L$) throughout the range of $eB$ considered, which is compatible with the observation from Fig.~\ref{kappa_ratio_static_wzB} where $\kappa_L$ curves dominate over $\kappa_T$. Even without the quark contribution, this dominance of $\kappa_L$ over $\kappa_T$ is hardly surprising because of the dominant gluonic contribution in the $t$-channel scatterings considered in the present study.

\section{Results II : Estimations beyond the static limit}
\label{sec7}

\begin{figure*}
\begin{center}
\includegraphics[scale=0.5]{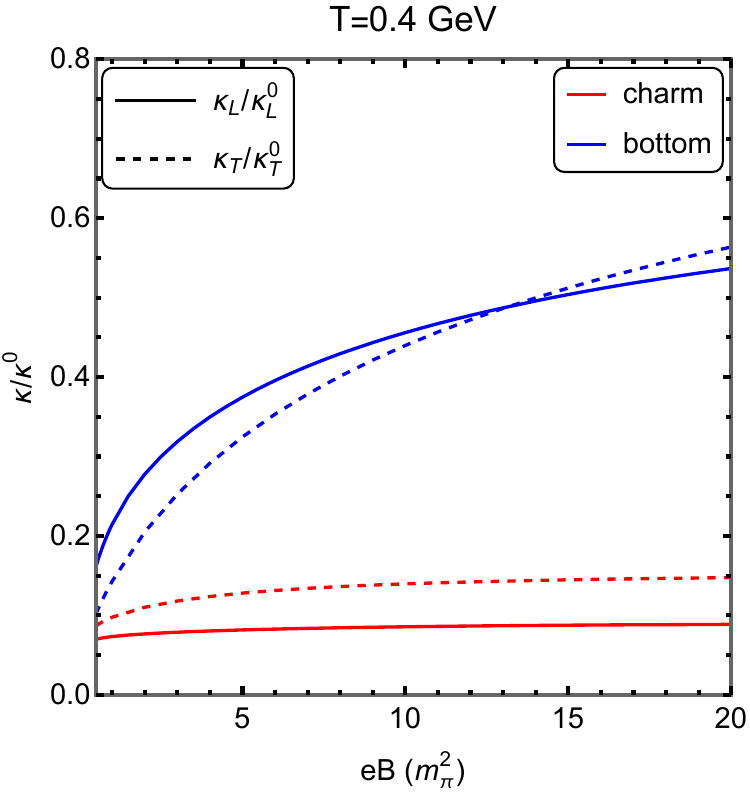}
\hspace{1cm}
\includegraphics[scale=0.5]{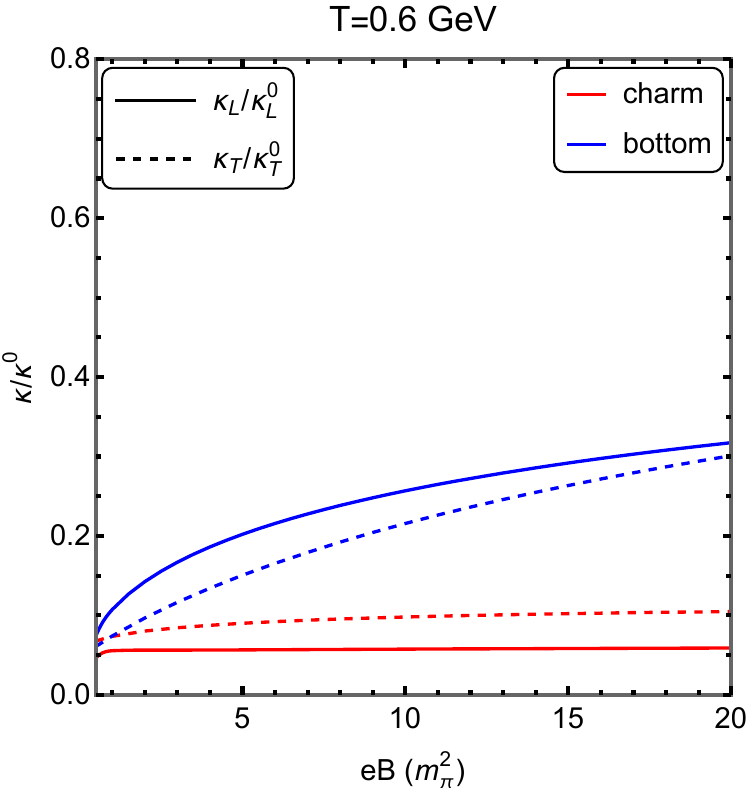}
\caption{$\vec{v}\sp \vec{B}$ case : Variation of the longitudinal (solid curves) and transverse (dashed curves) momentum diffusion coefficients for charm (red curves) and bottom (blue curves) quarks with external magnetic field for two different values of temperatures, i.e. $T=0.4$ GeV (left panel) and $T=0.6$ GeV (right panel). The magnetized momentum diffusion coefficients are scaled with respect to their $eB=0$ counterparts. Charm and bottom quark masses $M$ are specified in the text and HQ momentum $p$ is taken to be $1$ GeV. } 
\label{kappaveB_case1}
\end{center}
\end{figure*}

\begin{figure*}
\begin{center}
\includegraphics[scale=0.5]{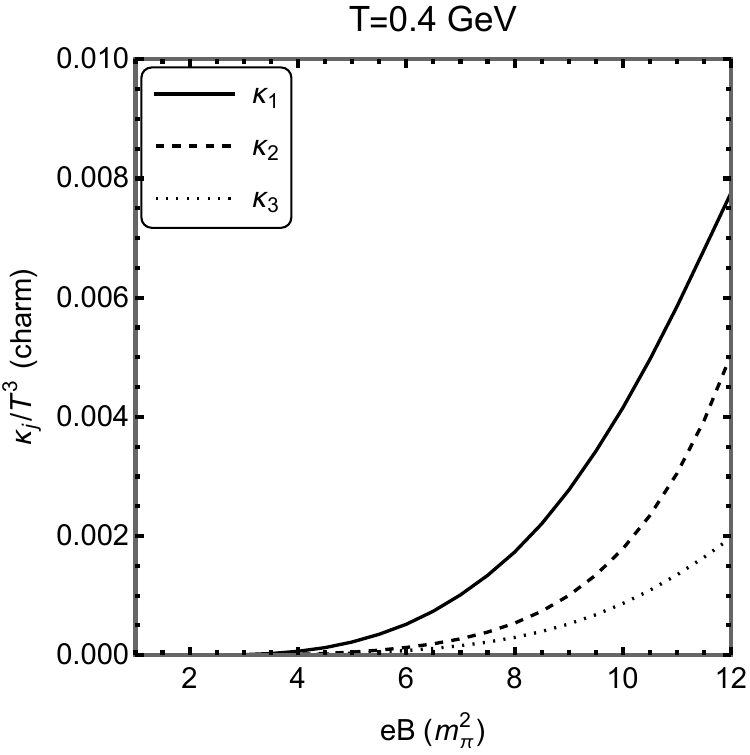}
\hspace{1cm}
\includegraphics[scale=0.5]{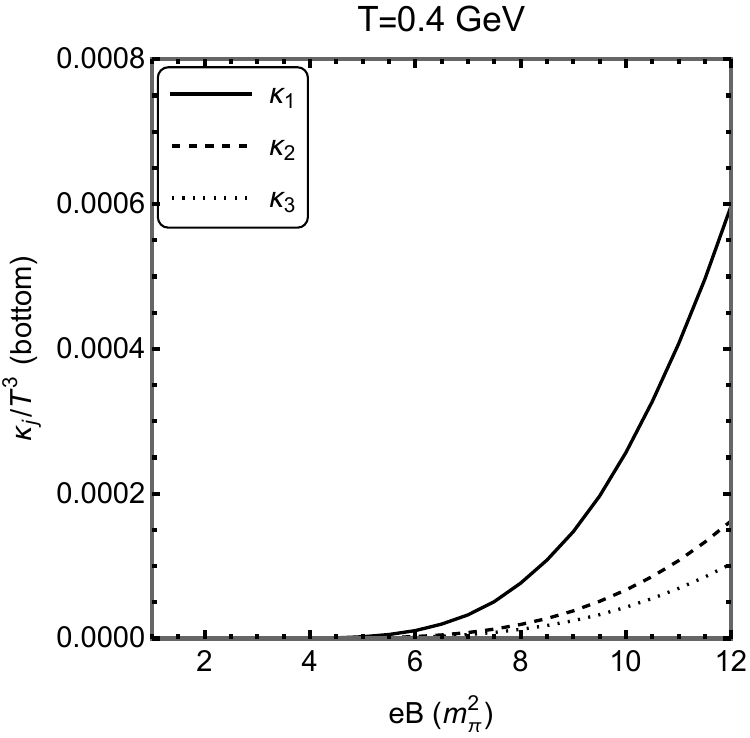}
\caption{$\vec{v}\perp \vec{B}$ case : Variation of the transverse components $\kappa_1$ (solid curves), $\kappa_2$ (dashed curves) and longitudinal component $\kappa_3$ (dotted curves) of the momentum diffusion coefficient for charm (left panel) and bottom (right panel) quarks with external magnetic field for a fixed value of temperature, i.e. $T=0.4$ GeV. The magnetized momentum diffusion coefficients are scaled with respect to their $eB=0$ counterparts. Charm and bottom quark masses $M$ are specified in the text and HQ momentum $p$ is taken to be $1$ GeV.} 
\label{kappaveB_case2}
\end{center}
\end{figure*}

In the previous section we have thoroughly analyzed our results within the static limit by scaling them with respect to the $eB=0$ result (Fig.~\ref{kappa_ratio_static_wzB}) and results from an approximated alternate procedure (Fig.~\ref{kappa_ratio_static}). In this section we show our estimated results for longitudinal and transverse momentum diffusion coefficients going beyond the static limit, but keeping ourselves confined within the limit of small energy transfer. Before discussing the results shown in this section, let us point out some necessary information about the parameters chosen for the current study.
\begin{enumerate}
    \item The form factors $d_i$s for the gluon two-point correlation function have been explicitly evaluated in Section~\ref{sec3}. Within the limit of small energy transfer, $d_i$s have further simplified expressions, discussed in Appendix~\ref{appC}. Those simplified $d_i$s have been further used to evaluate the spectral functions depicted in Appendix~\ref{appE}.
    
    \item The other factors $A_i$s, originating from the Dirac trace of the scattering rate are already $q_0$ independent, as explained in section~\ref{sec4}, so doesn't encounter further simplifications in the small energy transfer limit. Although for the two cases considered in the current study, i.e. $\vec{v}\sp \vec{B}$ and $\vec{v}\perp \vec{B}$, $A_i$'s obtain specific forms because of different vanishing components of HQ momentum $p$. Such forms of $A_i$s have been explicitly given in Appendix~\ref{appD}. 
    
    \item As also mentioned in the previous section for the static limit results, here also we use two kinds of one-loop running couplings, i.e. $g(T)$ and $g(T,eB)$. For the pure-glue contributions we use the normal temperature dependent coupling $g(T)$ whereas for the magnetized quark contributions we use the magnetic field modified running coupling $g(T,eB)$ from Refs.~\cite{Ayala:2014uua,Ayala:2016bbi,Ayala:2018wux,Ayala:2019nna}.
    
    \item The values of the charm and bottom quark masses are taken to be $M_c = 1.27$ GeV and $M_b = 4.18$ GeV respectively. Maintaining the scale hierarchy $M\ge p \gg T$, we have considered the value of the HQ momentum to be $p=1$ GeV and taken temperature values $T=0.4$ and $0.6$ GeV which also respect the applied HTL approximation. This in turn means that the $\vec{v}\sp \vec{B}$ case corresponds to $p_3=1$ GeV and $\vec{v}\perp \vec{B}$ case corresponds to $p_\perp = \sqrt{p_1^2+p_2^2} = 1$ GeV where for numerical simplicity we have chosen $p_2=0$ and $p_1=1$ GeV. 
    
    \item For the $\vec{v}\sp \vec{B}$ case, the exponential factor $e^{-\frac{q_\perp^2}{|q_f B|}}$ acts as a natural regulator for cutting off the UV divergences appearing from the $q$-integration, which also applies to our static limit results $\kappa_{L/T}^{(s)}$. On the other hand for the $\vec{v}\perp \vec{B}$ case we have a more general factor $e^{-\frac{(p_\perp-q_\perp)^2}{|q_f B|}}$ which is plagued by the huge value of the HQ momentum $p\approx 1$ GeV. Hence for this case obtaining UV-finite values require choosing a medium dependent cutoff similar to $q_{max}(T,eB)=3.1 T g(T,eB)^{1/3}$ chosen for the alternate procedure in section~\ref{sec6}.
    
    \item In the following results we have shown the variation of the momentum diffusion coefficients with respect to the external magnetic field. It showcases our flexibility of choosing any values of the magnetic field owing to the generality of our calculation. Due to the restrictions given on the lower values of temperature by HTL approximation, we refrain from showing the temperature variation throughout a certain range and instead opt to show different fixed values of temperature. Of course the effect of the magnetic field is more prominently reflected on the respective observable when varied with respect to $eB$ instead of $T$.  
\end{enumerate}

We are now in a position to describe the results. Let us start with the $\vec{v}\sp \vec{B}$ case, within which the variation of the longitudinal and transverse momentum diffusion coefficients with external magnetic field are shown in Fig.~\ref{kappaveB_case1} for two different fixed values of $T$, i.e. $0.4$ and $0.6$ GeV. The coefficients are scaled with their $eB=0$ values separately evaluated using the $eB=0$ scattering rate given in Eq.~\ref{sr_zeroB}. A similar trend like the static limit can be noticed here, for higher values of $eB$, the rate of change of $\kappa_{L/T}$ becomes rather flat, specially for the charm quarks (red curves). Interestingly for the charm quarks, $\kappa_T$ (dashed curves) dominates over $\kappa_L$ (solid curves) throughout the $eB$ range considered, in accordance with the trend observed in Ref.~\cite{Finazzo:2016mhm}. On the other hand for bottom quarks (blue curves), $\kappa_L$ is larger than $\kappa_T$ until a certain higher value of $eB$, after which we can notice a crossover. These crossover reflects the behaviours of three competing scales $M, T$ and $eB$. A much higher value of $M$ for the bottom quarks requires a much higher value of $eB$ to have similar behaviours of $\kappa_{L/T}$ for the charm quarks. A higher value of $T$ demands even more $eB$ which is evident from the right panel of Fig.~\ref{kappaveB_case1}, where no crossover is present but one can notice the converging trend of the bottom quark curves which predict a crossover at an even higher value of $eB$, than the values presented. 

Fig.~\ref{kappaveB_case2} describes the results from the other case, i.e. $\vec{v}\perp \vec{B}$. At $eB=0$, no such case can exist and hence we scaled the values of $\kappa_j$ with $T^3$ here. Transverse components $\kappa_1$ (solid curves) and $\kappa_2$ (dashed curves) dominates over the longitudinal component $\kappa_3$ (dotted curves) for both charm and bottom quarks, whereas $\kappa_1$ is larger than $\kappa_2$ because of the specific choice of $p_\perp$ considered. Although in this case, no saturating behaviours can be noticed for any of the momentum diffusion coefficients at higher values of $eB$ unlike the previous cases. Instead the rate of change of the momentum diffusion coefficients evidently increase with increasing values of $eB$. Also the values of momentum diffusion coefficients for the bottom quarks (right panel) are an order of magnitude lower than that for the charm quakrs (left panel).

Combining the results from the two different cases, one can observe that for lower values of $eB$ the effect of the external magnetic field on various momentum diffusion coefficients is higher when the HQ moves along the direction of the magnetic field with considerably high momentum. This effect saturates when we increase the value of $eB$. On the other hand when the HQ moves in the perpendicular direction of the external magnetic field, the effect of the magnetic field on the momentum diffusion coefficients increases with increasing values of $eB$.

\section{Summary}
\label{sec8}

In the present study we have studied HQ momentum diffusion coefficients in magnetized medium. We have considered the most general cases in both the fronts : (a) arbitrary values of the external magnetic field and (b) going beyond the static limit of the HQ to consider a finite HQ velocity. As far as we know, this is the first calculation which attempts to study the HQ dynamics with arbitrary values of the external magnetic field, i.e. incorporates the contributions from higher Landau levels. Hence the novelty of this work mainly lies in the calculation part, summarized in section~\ref{sec2} and subsequently performed in sections~\ref{sec3}, \ref{sec4} along with the various appendices. 

We have also discussed an alternate procedure, usually employed in evaluating other HQ observable in a magnetized medium, which assumes the whole medium effect within the Debye screening mass and accordingly replace it with the magnetized medium modified one. By comparing the results from this approximated procedure with our exact results within the static limit of the HQ, we clearly identify the shortcomings of this alternate procedure and emphasise the importance of employing the general structure of the gluon two-point correlation functions in a hot and arbitrarily magnetized medium. Both within and beyond (when we consider the HQ to be moving along the direction of the external magnetic field) the static limit we notice similar pattern in the $eB$ dependence of the momentum diffusion coefficients, i.e. the $eB$ dependence is rapidly increasing for lower values of $eB$, whereas it becomes saturated (more so for charm quarks) for relatively higher values of $eB$. An opposite trend is observed for the case when the HQ is moving in a perpendicular direction with respect to the external magnetic field, where increasing $eB$ dictates more changes in the momentum diffusion coefficients. Soft gluon scattering (which are dominant within the $t$-channel at leading order of strong coupling) governed longitudinal diffusion coefficient dominates over the transverse diffusion coefficient within the static limit of HQ and an opposite trend is observed going beyond the static limit. Competing behaviour of various scales involved, i.e. $M$, $p$, $eB$ and $T$, gets well reflected in our results.  

At this point we must also mention that the present work is also not completely free from limitations, most of which are being carried over from the limitations of HTL approximation, e.g. assuming the quarks in various Landau levels to be mass less renders into vanishing quark contribution to the longitudinal momentum diffusion coefficient. At the cost of providing completely analytic, gauge independent, simplified expressions there are further scale restrictions appearing because of the HTL approximation, as discussed in section~\ref{sec1}. But apart from the HTL generated limitations, there is another important issue of the UV cutoff $q_{max}$. For $eB=0$, several studies have shown that the UV cutoff is not necessary when one includes the hard contributions along with the soft contribution in the scattering rate. In most of the cases with finite $eB$, the exponential factor $e^{-q_\perp^2/|q_f B|}$ acts as a natural UV cutoff. But for the case of HQ moving with sufficiently high transverse momentum the exponential factor is not enough and we need further assurances in the form of a medium dependent $q_{max}(T,eB)$. In the present work, this $q_{max}(T,eB)$ has been chosen by choosing a similar form as the $eB=0$ case evaluated in Ref~\cite{Beraudo:2009pe} by comparing with the full result. But this choice of $q_{max}$ is not completely accurate and requires further modifications. One of the way of modifying this $q_{max}$ is also discussed in the present work, which requires exact comparison between two alternate procedures within the HQ static limit and extracting the $q_{max}$ by appropriate fitting procedure. But the ultimate way of eradicating this $q_{max}$ dependence would be proper inclusion of the hard contributions~\cite{Braaten:1991jj,Braaten:1991we}, which leaves the door open for potential future investigations. The present study creates other avenues to proceed too, e.g. examining the HQ in-medium evolution using a Langevin transport code (e.g.~\cite{Li:2019lex}) and their consequences on the experimental observable such as directed and elliptic flow of the open heavy flavor mesons.

\begin{acknowledgments}
This work was supported by the postdoctoral research fellowship of Alexander von Humboldt Foundation.
\end{acknowledgments}

\appendix


\section{Trace for one-loop gluon self energy}
\label{appA}

Here we note down the explicit expressions for the individual terms $(T_i)_{\mn}$, given in Eq.~\eqref{trace_total}.
\begin{align}
T_1 &= L_l L_{l'}\textsf{Tr}\Bigl[\gamma_\mu \slashed{K}_\sp\left(1-i\gamma_1\gamma_2\right)\gamma_\nu\slashed{R}_\sp\left(1-i\gamma_1\gamma_2\right)\Bigr]\nn\\
&= 8 L_l L_{l'}(K^\sp_\mu R^\sp_\nu + K^\sp_\nu R^\sp_\mu - g_{\mn}^\sp (K\cdot R)_\sp) \\
T_2 &= L_{l-1}L_{l'-1} \textsf{Tr}\Bigl[\gamma_\mu \slashed{K}_\sp\left(1+i\gamma_1\gamma_2\right)\gamma_\nu\slashed{R}_\sp\left(1+i\gamma_1\gamma_2\right)\Bigr]\nn\\
&= 8 L_{l-1}L_{l'-1}(K^\sp_\mu R^\sp_\nu +K^\sp_\nu R^\sp_\mu - g_{\mn}^\sp (K\cdot R)_\sp) \\
T_3 &= -L_{l}L_{l'-1}\textsf{Tr}\Bigl[\gamma_\mu \slashed{K}_\sp\left(1-i\gamma_1\gamma_2\right)\gamma_\nu\slashed{R}_\sp\left(1+i\gamma_1\gamma_2\right)\Bigr]\nn\\
&= -8 L_l L_{l'-1}(g_{\mn}^\perp + i (g_{2\mu}g_{1\nu} - g_{1\mu}g_{2\nu}))(K\cdot R)_\sp \\
T_4 &= -L_{l-1}L_{l'} \textsf{Tr}\Bigl[\gamma_\mu \slashed{K}_\sp\left(1+i\gamma_1\gamma_2\right)\gamma_\nu\slashed{R}_\sp\left(1-i\gamma_1\gamma_2\right)\Bigr]\nn\\
&= -8 L_{l-1}L_{l'}(g_{\mn}^\perp - i (g_{2\mu}g_{1\nu} - g_{1\mu}g_{2\nu}))(K\cdot R)_\sp \\
T_5 &= 16 L_{l-1}^1 L_{l'-1}^1\textsf{Tr}\Bigl[\gamma_\mu\slashed{K}_\perp\gamma_\nu\slashed{R}_\perp\Bigr]\nn\\
&= 64L_{l-1}^1L_{l'-1}^1(K^\perp_\mu R^\perp_\nu + K^\perp_\nu R^\perp_\mu + g_{\mn} (k\cdot r)_\perp) \\
T_6 &= -4L_lL_{l'-1}^1\textsf{Tr}\Bigl[\gamma_\mu \slashed{K}_\sp\left(1-i\gamma_1\gamma_2\right)\gamma_\nu\slashed{R}_\perp\Bigr]\nn\\
&= -16L_l L_{l'-1}^1[K_\mu^\sp R_\nu^\perp + K_\nu^\sp R_\mu^\perp - ir_1(K_\mu^\sp g_{2\nu} - K_\nu^\sp g_{2\mu})\nn\\
& ~~- ir_2(K_\nu^\sp g_{1\mu} - K_\mu^\sp g_{1\nu})] \\
T_7 &= 4L_{l-1}L_{l'-1}^1 \textsf{Tr}\Bigl[\gamma_\mu \slashed{K}_\sp\left(1+i\gamma_1\gamma_2\right)\gamma_\nu\slashed{K}_\perp\Bigr]\nn\\
&= 16L_{l-1}L_{l'-1}^1 [K_\mu^\sp R_\nu^\perp + K_\nu^\sp R_\mu^\perp  + ir_1(K_\mu^\sp g_{2\nu} - K_\nu^\sp g_{2\mu}) \nn\\
&~~+ ir_2(K_\nu^\sp g_{1\mu} - K_\mu^\sp g_{1\nu})] \\
T_8 &= -4L_{l-1}^1 L_{l'}\textsf{Tr}\Bigl[\gamma_\mu \slashed{K}_\perp\gamma_\nu\slashed{R}_\sp\left(1-i\gamma_1\gamma_2\right)\Bigr]\nn\\
&= -16L_{l-1}^1 L_{l'} [K_\nu^\perp R_\mu^\sp  + K_\mu^\perp R_\nu^\sp + ik_1(R_\mu^\sp g_{2\nu} - R_\nu^\sp g_{2\mu})\nn\\
&~~+ ik_2(R_\nu^\sp g_{1\mu} - R_\mu^\sp g_{1\nu})] \\
T_9 &= 4L_{l-1}^1 L_{l'-1}\textsf{Tr}\Bigl[\gamma_\mu \slashed{K}_\perp\gamma_\nu \slashed{R}_\sp\left(1+i\gamma_1\gamma_2\right)\Bigr]\nn\\
&= 16L_{l-1}^1 L_{l'-1} [K_\nu^\perp R_\mu^\sp  + K_\mu^\perp R_\nu^\sp - ik_1(R_\mu^\sp g_{2\nu} - R_\nu^\sp g_{2\mu}) \nn\\
&~~ - ik_2(R_\nu^\sp g_{1\mu} - R_\mu^\sp g_{1\nu})] 
\end{align}

Here for brevity we have written $L_l(\xi_k^\perp)$ and $L_{l'}(\xi_r^\perp)$ as $L_l$ and $L_{l'}$ etc. Also the terms $(g_{2\mu}g_{1\nu} - g_{1\mu}g_{2\nu})$ can be written in terms of the electromagnetic field tensor $F_{\mn}$~\cite{Wang:2021ebh}.

\section{Contractions and simplifications required for the evaluation of the form factors}
\label{appB}

\subsection{Form factor $d_1^m$}
For $d_1^m$, i.e. the contribution from the quark loop, below we note down the contractions coming from the $(T_i)_{\mn}$ part : 
\begin{align}
    \Delta_1^{\mn} (T_1 +T_2) &= \frac{8}{\bar{u}^2} (L_l L_{l'} +L_{l-1}L_{l'-1}) \nn\\
    &\left[2(\bar{u}\cdot K)_\sp (\bar{u}\cdot R)_\sp - \bar{u}_\sp^2 (K\cdot R)_\sp\right]\\
    \Delta_1^{\mn} (T_3 +T_4) &= \frac{8\bar{u}_\perp^2}{\bar{u}^2} (L_l L_{l'-1} +L_{l-1}L_{l'})(K\cdot R)_\sp\\
    \Delta_1^{\mn}T_5 &= \frac{64}{\bar{u}^2} L_{l-1}^1L_{l'-1}^1 \nn\\
    &\left[2(\bar{u}\cdot k)_\perp (\bar{u}\cdot r)_\perp +\bar{u}^2 (k\cdot r)_\perp \right]\\
    \Delta_1^{\mn} (T_6 +T_7) &= \frac{32}{\bar{u}^2} (L_{l}L_{l'-1}^1 - L_{l-1}L_{l'-1}^1) \nn\\
    &\left[(\bar{u}\cdot K)_\sp (\bar{u}\cdot r)_\perp\right]\\
    \Delta_1^{\mn} (T_8 +T_9) &= \frac{32}{\bar{u}^2} (L_{l-1}^1 L_{l'} - L_{l-1}^1 L_{l'-1}) \nn\\
    &\left[(\bar{u}\cdot k)_\perp (\bar{u}\cdot R)_\sp\right]
\end{align}

Now within the HTL approximation we can neglect the soft external momenta $Q$ in the numerator and approximate $K\approx R$. These approximations further simplifies the $\Delta_1^{\mn}T_{\mn}$ as -
\begin{align}
    \Delta_1^{\mn}T_{\mn} &\approx \frac{8}{\bar{u}^2} (L_l L_{l'} +L_{l-1}L_{l'-1}) (2k_0^2 - \bar{u}_\sp^2 K_\sp^2) \nn\\
    & +\frac{8\bar{u}_\perp^2}{\bar{u}^2} (L_l L_{l'-1} +L_{l-1}L_{l'}) K_\sp^2 \nn\\
    & + 64 L_{l-1}^1L_{l'-1}^1 k_\perp^2.
\end{align}

HTL approximation also allows us to simplify the perpendicular momentum integration using the following identities 
\begin{align}
    &\int\frac{d^2k_\perp}{(2\pi)^2}\exp\left(-\frac{2k_\perp^2}{|q_fB|}\right)L_l(\xi_k^\perp) L_{l'}(\xi_k^\perp)\nn\\
    &= \frac{|q_f B|}{8\pi}\delta_{l,l'},\\
    &\int\frac{d^2k_\perp}{(2\pi)^2}\exp\left(-\frac{2k_\perp^2}{|q_fB|}\right)k_\perp^2~L_l^1(\xi_k^\perp) L_{l'}^1(\xi_k^\perp)\nn\\
    &= \frac{|q_fB|^2}{16\pi}(l+1)\delta_{l+1,l'+1}.
\end{align}

Incorporating all these we can finally write down the expression for $d_1^m$, as given in Eq.~\eqref{coeff_d1m}.

\subsection{Form factor $d_2^m$}
For the contribution $d_2^m$, we again list down the contractions $\Delta_2^{\mn} (T_i)_{\mn}$ first. 
\begin{align}
&\Delta_2^{\mn} (T_1+T_2+T_6+T_7+T_8+T_9) =0 \\
&\Delta_2^{\mn} (T_3+T_4) = -8 (L_{l}L_{l'-1}+L_{l-1}L_{l'}) (K\cdot R)_\sp \\
&\Delta_2^{\mn} T_5 = 64L_{l-1}^1L_{l'-1}^1\left\{2\frac{(q\cdot k)_\perp(q\cdot r)_\perp}{q_\perp^2} -(q\cdot r)_\perp\right\}
\end{align}

So, within the HTL approximation, we can write down the final contraction as 
\begin{align}
    \Delta_2^{\mn} T_{\mn} &\approx -8 (L_{l}L_{l'-1}+L_{l-1}L_{l'}) K_\sp^2 \nn\\
    &~~- 64L_{l-1}^1L_{l'-1}^1 k_\perp^2.
\end{align}

Using similar techniques for the perpendicular momentum integration and the frequency sums $\Phi_1, \Phi_2$ we can readily obtain the final expression for $d_2^m$, as given in Eq.~\eqref{coeff_d2m}.

\subsection{Form factor $d_3^m$}

For $d_3^m$, first we note down the contractions coming from the $(T_i)_{\mn}$ part :  
\begin{align}
    \Delta_3^{\mn} (T_1 +T_2) &= \frac{8}{\bar{n}^2} (L_l L_{l'} +L_{l-1}L_{l'-1}) \nn\\
    &\left[2(\bar{n}\cdot K)_\sp (\bar{n}\cdot R)_\sp - \bar{n}_\sp^2 (K\cdot R)_\sp\right]\\
    \Delta_3^{\mn} (T_3 +T_4) &= \frac{8\bar{n}_\perp^2}{\bar{n}^2} (L_l L_{l'-1} +L_{l-1}L_{l'})(K\cdot R)_\sp\\
    \Delta_3^{\mn}T_5 &= \frac{64}{\bar{n}^2} L_{l-1}^1L_{l'-1}^1 \nn\\
    &\left[2(\bar{n}\cdot k)_\perp (\bar{n}\cdot r)_\perp +\bar{n}^2 (k\cdot r)_\perp \right]\\
    \Delta_3^{\mn} (T_6 +T_7) &= \frac{32}{\bar{n}^2} (L_{l}L_{l'-1}^1 - L_{l-1}L_{l'-1}^1) \nn\\
    &\left[(\bar{n}\cdot K)_\sp (\bar{n}\cdot r)_\perp\right]\\
    \Delta_3^{\mn} (T_8 +T_9) &= \frac{32}{\bar{n}^2} (L_{l-1}^1 L_{l'} - L_{l-1}^1 L_{l'-1}) \nn\\
    &\left[(\bar{n}\cdot k)_\perp (\bar{n}\cdot R)_\sp\right]
\end{align}

Now within the HTL approximation further simplifications yields -
\begin{align}
    \Delta_3^{\mn}T_{\mn} &\approx -\frac{8\bar{n}_\sp^2}{\bar{n}^2} (L_l L_{l'} +L_{l-1}L_{l'-1}) (2k_3^2 + K_\sp^2) \nn\\
    & +\frac{8\bar{n}_\perp^2}{\bar{n}^2} (L_l L_{l'-1} +L_{l-1}L_{l'}) K_\sp^2 \nn\\
    & + 64 L_{l-1}^1L_{l'-1}^1 k_\perp^2
\end{align}

Subsequently after performing the perpendicular momentum integration and in terms of the frequency sums we can write down the final expression of $d_3^m$ as Eq.~\eqref{coeff_d3m}.

\subsection{Form Factor $d_4$}

Finally for $d_4^m$, again we start with listing the contractions, 
\begin{align}
    &\Delta_4^{\mn} (T_1 +T_2) = \frac{16}{\sqrt{\bar{u}^2}\sqrt{\bar{n}^2}} (L_l L_{l'} +L_{l-1}L_{l'-1}) \nn\\
    &\left[(\bar{u}\cdot K)_\sp (\bar{n}\cdot R)_\sp + (\bar{n}\cdot K)_\sp (\bar{u}\cdot R)_\sp - (\bar{n}\cdot\bar{u})_\sp (K\cdot R)_\sp\right]\\
    &\Delta_4^{\mn} (T_3 +T_4) = \frac{16(\bar{n}\cdot\bar{u})_\perp}{\sqrt{\bar{u}^2}\sqrt{\bar{n}^2}} (L_l L_{l'-1} +L_{l-1}L_{l'})(K\cdot R)_\sp\\
    &\Delta_4^{\mn}T_5 = \frac{128}{\sqrt{\bar{u}^2}\sqrt{\bar{n}^2}} L_{l-1}^1L_{l'-1}^1 \left[(\bar{u}\cdot k)_\perp (\bar{n}\cdot r)_\perp + \right. \nn\\
    &\left. (\bar{n}\cdot k)_\perp (\bar{u}\cdot r)_\perp + (\bar{n}\cdot\bar{u}) (k\cdot r)_\perp \right]\\
    &\Delta_4^{\mn} (T_6 +T_7) = \frac{32}{\sqrt{\bar{u}^2}\sqrt{\bar{n}^2}} (L_{l}L_{l'-1}^1 - L_{l-1}L_{l'-1}^1) \nn\\
    &\left[(\bar{u}\cdot K)_\sp (\bar{n}\cdot r)_\perp + (\bar{n}\cdot K)_\sp (\bar{u}\cdot r)_\perp\right]\\
    &\Delta_4^{\mn} (T_8 +T_9) = \frac{32}{\sqrt{\bar{u}^2}\sqrt{\bar{n}^2}} (L_{l-1}^1 L_{l'} - L_{l-1}^1 L_{l'-1}) \nn\\
    &\left[(\bar{u}\cdot k)_\perp (\bar{n}\cdot R)_\sp + (\bar{n}\cdot k)_\perp (\bar{u}\cdot R)_\sp\right].
\end{align}

Within the HTL approximation, the simplified version of the contraction can be expressed as -
\begin{align}
    &\Delta_4^{\mn}T_{\mn} \approx \frac{16}{\sqrt{\bar{u}^2}\sqrt{\bar{n}^2}} (L_l L_{l'} +L_{l-1}L_{l'-1}) (2k_0 k_3\bar{n}^2\nn\\
    &~~- (\bar{n}\cdot\bar{u})_\sp K_\sp^2) +\frac{16(\bar{n}\cdot\bar{u})_\perp}{\sqrt{\bar{u}^2}\sqrt{\bar{n}^2}} (L_l L_{l'-1} +L_{l-1}L_{l'}) K_\sp^2 
\end{align}

After performing the perpendicular momentum integration, the final expression of $d_4^m$ can be written as given in Eq.~\eqref{coeff_d4m}.


\section{Small energy transfer limit and further simplifications of form factors}
\label{appC}

In the limit of small energy transfer, we can assume $q_0\rightarrow 0$, which results in further simplifications of the form factors. In this limit we can assume $\bar{u}^2 = \bar{u}_\sp^2 = 1$ and $\bar u_\perp^2= (\bar n \cdot \bar u)_\sp = (\bar n \cdot \bar u)_\perp = 0$. Below we write down the final expressions of $d_i$'s within small energy transfer limit, which has been used further in our computation of spectral functions. We also absorb the extra sum over Landau levels using the Kronecker delta functions.  

\begin{align}
d_1 &=\frac{N_cg^2T^2}{3}\left[1-\mathcal{T}_Q\right] -\sum_f~\frac{g^2 |q_fB|}{2\pi}\sum_{l=0}^\infty\nn\\
&\left[(2-\delta_{l,0})\int\frac{dk_3}{2\pi}\left\{\Phi_1+2(k_3^2+ 2l|q_fB|)\Phi_2\Big|_{l'=l}\right\}\right],\label{coeff_d1_final}
\end{align}

\begin{align}
d_2 &=\frac{N_cg^2T^2}{6}\mathcal{T}_Q + \sum_f~\frac{g^2 |q_fB|}{2\pi}\sum_{l=1}^\infty\left[(-1)\right. \nn\\
&\left.\int\frac{dk_3}{2\pi}\left\{2\Phi_1+2l|q_fB|\left(\Phi_2\Big|_{l'=l+1}+\Phi_2\Big|_{l'=l-1}\right)\right\}\right.\nn\\
&\left.+4|q_fB|l\int\frac{dk_3}{2\pi}\Phi_2\Big|_{l'=l}\right],
\label{coeff_d2_final}
\end{align}

\begin{align}
d_3 &= \frac{N_cg^2T^2}{6}\mathcal{T}_Q -\sum_f~\frac{g^2 |q_fB|}{2\pi\bar{n}^2}\sum_{l=0}^\infty \left[-\bar{n}_\sp^2(2-\delta_{l,0}) \right.\nn\\
&\int\frac{dk_3}{2\pi}\left\{\Phi_1+(2k_3^2+2l|q_fB|)\Phi_2\Big|_{l'=l}\right\}-(1-\delta_{l,0})\bar{n}_\perp^2\nn\\
&\int\frac{dk_3}{2\pi}\left\{2\Phi_1+2l|q_fB|\left(\Phi_2\Big|_{l'=l+1}+\Phi_2\Big|_{l'=l-1}\right)\right\}\nn\\
&\left.+4\bar{n}^2|q_fB|l\int\frac{dk_3}{2\pi}\Phi_2\Big|_{l'=l}\right],
\label{coeff_d3_final}
\end{align}

\begin{align}
d_4 &= -\sum_f~\frac{g^2 |q_fB|}{2\pi\sqrt{\bar{n}^2}}\sum_{l=0}^\infty\left[2\bar{n}^2(2-\delta_{l,0})\int\frac{dk_3}{2\pi} k_3\Phi_3\Big|_{l'=l}\right].
\label{coeff_d4_final}
\end{align}

We can further simplify the frequency sum $\Phi_i$'s within the limit of small energy transfer and employing the specific conditions between the Landau levels $l$ and $l'$. We deal with them one by one. First of all, at $l=l'$, we can approximate $E_{r_3}\approx E_{k_3} - q_3$ and hence $n_F(E_{r_3})\approx n_F(E_{k_3})-q_3\frac{\partial n_F(E_{k_3})}{\partial E_{k_3}}$. Subsequently we can simplify the frequency sums as
\begin{align}
\Phi_2\Big|_{l'=l} &= -\sum_{s_1,s_2=\pm 1}\frac{s_1s_2}{4E_{k_3}^2}\left(\frac{1-n_F(s_1E_{k_3})-n_F(s_2E_{r_3})}{q_0-s_1E_{k_3}-s_2E_{r_3}}\right) \nn\\
&= -\frac{n_F(E_{k_3})}{2E_{k_3}^3}-\frac{q_3}{4E_{k_3}^2}\frac{\partial n_F(E_{k_3})}{\partial E_{k_3}}\times\nn\\
&~~~~~ \left[\frac{1}{q_0-q_3} -\frac{1}{q_0+q_3} \right],\label{Phi2_equal_ll} \\
\Phi_3\Big|_{l'=l} &= -\sum_{s_1,s_2=\pm 1}\frac{s_2}{4E_{k_3}}\left(\frac{1-n_F(s_1E_{k_3})-n_F(s_2E_{r_3})}{q_0-s_1E_{k_3}-s_2E_{r_3}}\right) \nn\\
&=-\frac{q_3}{4E_{k_3}}\frac{\partial n_F(E_{k_3})}{\partial E_{k_3}}\left[\frac{1}{q_0-q_3} +\frac{1}{q_0+q_3} \right].
\end{align}
Moreover one can clearly notice that in the limit of small energy transfer, i.e. $q_0\rightarrow 0$, $\Phi_3\Big|_{l'=l} \approx 0$, which in turn suggests $d_4\approx 0$. Other relevant frequency sums, i.e. $\Phi_2\Big|_{l'=l+1},~\Phi_2\Big|_{l'=l-1}$, will not go through further simplification process and can be expressed by simply replacing $l'=l\pm 1$ respectively within $E_{r_3}$ in Eq.~\eqref{phi2_initial}.

\section{Evaluation of $A_i$'s}
\label{appD}

In this section we will evaluate the individual traces posed in Eq.~\eqref{traces_Ai}. First we will obtain the most general expressions for $A_i$'s and then discuss the special cases one by one. We provide the result for each of the traces as follows ($K=P-Q$):

  \begin{align}
&\Tr\left[(\slashed{P}+M)\Delta_1^{\mn}\gamma_\mu D_l(q_fB,K)\gamma_\nu\right]\nn\\
 &~~= \frac{4}{\bar{u}^2} \left[2(\bar{u}\cdot K)_\sp (\bar{u}\cdot P)-\bar{u}^2\left((K\cdot P)_\sp-M^2\right)\right]\times \nn\\
 & (L_l\left(\xi_k^\perp\right) - L_{l-1}\left(\xi_k^\perp\right)) +\frac{16}{\bar{u}^2}[2(\bar{u}\cdot k)_\perp(\bar{u}\cdot P)\nn\\
 &-\bar{u}^2(k\cdot p)_\perp] L_{l-1}^1\left(\xi_k^\perp\right) \nn\\
&~~=A_1 + B_1(q_0),\label{traces_A1_done}\\
&\Tr\left[(\slashed{P}+M)\Delta_2^{\mn}\gamma_\mu D_l(q_fB,K)\gamma_\nu\right]\nn\\
&~~= 4\left[M^2-(K\cdot P)_\sp\right](L_l\left(\xi_k^\perp\right) - L_{l-1}\left(\xi_k^\perp\right)) \nn\\
&~~-16\left[2\frac{(q\cdot k)_\perp(q\cdot p)_\perp}{q_\perp^2} -(k \cdot p)_\perp\right]L_{l-1}^1\left(\xi_k^\perp\right)\nn\\
&~~= A_2 + B_2(q_0), \label{traces_A2_done} 
\end{align}

\begin{align}
&\Tr\left[(\slashed{P}+M)\Delta_3^{\mn}\gamma_\mu D_l(q_fB,K)\gamma_\nu\right]\nn\\
 =& \frac{4}{\bar{n}^2} \left[2(\bar{n}\cdot K)_\sp (\bar{n}\cdot P)-\bar{n}^2\left((K\cdot P)_\sp-M^2\right)\right]\times \nn\\
 & (L_l\left(\xi_k^\perp\right) - L_{l-1}\left(\xi_k^\perp\right)) +\frac{16}{\bar{n}^2}[2(\bar{n}\cdot k)_\perp(\bar{n}\cdot P)\nn\\
 &~~-\bar{n}^2(k\cdot p)_\perp] L_{l-1}^1\left(\xi_k^\perp\right) \nn\\
 =&A_3 + B_3(q_0),\label{traces_A3_done}\\
&\Tr\left[(\slashed{P}+M)\Delta_4^{\mn}\gamma_\mu D_l(q_fB,K)\gamma_\nu\right]\nn\\
&= \frac{4}{\sqrt{\bar{u}^2}\sqrt{\bar{n}^2}} \big[(\bar{u}\cdot K)_\sp (\bar{n}\cdot P)+(\bar{n}\cdot K)_\sp (\bar{u}\cdot P)\nn\\
&-2(\bar{n}\cdot\bar{u})\left((K\cdot P)_\sp-M^2\right)\big] (L_l\left(\xi_k^\perp\right) - L_{l-1}\left(\xi_k^\perp\right)) \nn\\
& +\frac{32}{\sqrt{\bar{u}^2}\sqrt{\bar{n}^2}}[(\bar{n}\cdot K)_\perp(\bar{u}\cdot P)+(\bar{u}\cdot K)_\perp(\bar{n}\cdot P)] L_{l-1}^1\left(\xi_k^\perp\right) \nn\\
 &= A_4 + B_4(q_0),\label{traces_A4_done}
\end{align}  
 
From Eqs.~\eqref{traces_A1_done},\eqref{traces_A2_done},\eqref{traces_A3_done} and \eqref{traces_A4_done}, we can subsequently extract the $q_0$ independent $A_i$'s, given as
\begin{align}
A_1 &= \frac{4}{\bar{u}^2} \Bigl[2p_0^2+\bar{u}^2\left(M^2 - P_\sp^2 - p_3q_3\right)\Bigr](L_l - L_{l-1})\nn\\
& - 16 (k\cdot p)_\perp L^1_{l-1}.
\label{A1_final}\\
A_2 &= 4\left(M^2 - P_\sp^2 - p_3q_3\right)(L_l - L_{l-1})\nn\\
&~~-16\left(2\frac{(q\cdot p)_\perp^2}{q_\perp^2} - (q\cdot p)_\perp - p_\perp^2\right)L_{l-1}^1.
\label{A2_final}\\
A_3 &= 4\Bigl[\frac{2k_3q_3}{q^2}(\vec{p}\cdot\vec{q})+M^2 - p_0^2- p_3k_3\Bigr](L_l - L_{l-1})\nn\\
&+\frac{16}{\bar{n}^2}[2\frac{q_3(q\cdot k)_\perp}{q^2}\left(\frac{q_3(\vec{p}\cdot \vec{q})}{q^2}-p_3\right)\nn\\
&~~+\bar{n}^2\left((p\cdot q)_\perp-p_\perp^2\right)] L_{l-1}^1.
\label{A3_final}\\
A_4 & = \frac{4p_0}{\sqrt{\bar{u}^2}\sqrt{\bar{n}^2}} \big[\left(-p_3+\frac{q_3}{q^2}(\vec{p}\cdot\vec{q})\right)+k_3\bar{n}^2\big](L_l - L_{l-1})\nn\\
& +\frac{32}{\sqrt{\bar{u}^2}\sqrt{\bar{n}^2}} \frac{p_0q_3}{q^2}(q\cdot k)_\perp L_{l-1}^1.
\label{A4_final}
\end{align}

\subsection{case 1 : $\vec{v} \sp \vec{B}$}

For this case, we have $p_1=p_2=0$ and incorporating the small energy transfer limit of $\omega\rightarrow 0$, $A_i$'s will have the following form. 
\begin{align}
A_1^{(1)} &= 4 \Bigl[p_0^2+ M^2 + p_3^2 - p_3q_3\Bigr](L_l - L_{l-1}),
\label{A1_case1}\\
A_2^{(1)} &= 4\left(M^2 - p_0^2 +p_3^2 - p_3q_3\right)(L_l - L_{l-1}),
\label{A2_case1}\\
A_3^{(1)} &= 4\Bigl[k_3p_3(2\bar{n}^2+1)+M^2 - p_0^2\Bigr](L_l - L_{l-1})\nn\\
&+32p_3q_3\bar{n}^2 L_{l-1}^1,
\label{A3_case1}\\
A_4^{(1)} & = \frac{4p_0}{\sqrt{\bar{n}^2}} \big[(2p_3-q_3)\bar{n}^2\big](L_l - L_{l-1})\nn\\
& +\frac{32}{\sqrt{\bar{n}^2}} \frac{p_0q_3}{q^2}q_\perp^2 L_{l-1}^1.
\label{A4_case1}
\end{align}

\subsection{case 2 : $\vec{v} \perp \vec{B}$}

For this case, we have $p_3=0$ and incorporating the small energy transfer limit of $\omega\rightarrow 0$, $A_i$'s will have the following form. 
\begin{align}
A_1^{(2)} &= 4(p_0^2+ M^2)(L_l - L_{l-1}) - 16 (k\cdot p)_\perp L^1_{l-1},
\label{A1_case2}\\
A_2^{(2)} &= 4\left(M^2 - p_0^2\right)(L_l - L_{l-1})\nn\\
&~~-16\left(2\frac{(q\cdot p)_\perp^2}{q_\perp^2} - (q\cdot p)_\perp - p_\perp^2\right)L_{l-1}^1.
\label{A2_case2}\\
A_3^{(2)} &= 4\Bigl[-\frac{2q_3^2}{q^2}(p\cdot q)_\perp+M^2 - p_0^2\Bigr](L_l - L_{l-1})\nn\\
&+\frac{16}{\bar{n}^2}\left[2\frac{q_3^2(q\cdot k)_\perp(p\cdot q)_\perp}{q^4}+\bar{n}^2\left((p\cdot q)_\perp-p_\perp^2\right)\right] L_{l-1}^1.
\label{A3_case2}\\
A_4^{(2)} & = \frac{4p_0}{\sqrt{\bar{n}^2}} \left[\frac{q_3}{q^2}(p\cdot q)_\perp-q_3\bar{n}^2\right](L_l - L_{l-1})\nn\\
& +\frac{32}{\sqrt{\bar{n}^2}} \frac{p_0q_3}{q^2}(q\cdot k)_\perp L_{l-1}^1.
\label{A4_case2}
\end{align}

\subsection{static limit}

Within the static limit, we have ${\bf v} \rightarrow 0$, hence $p_1=p_2=p_3=0$ and $p_0\approx M$. So, within static limit $A_i$'s take the following form. 
\begin{align}
A_1^{(s)} &= 8M^2(L_l - L_{l-1}),\label{A1_static}\\
A_2^{(s)} &= 0,\label{A2_static}\\
A_3^{(s)} &= 0,\label{A3_static}\\
A_4^{(s)} & = -\frac{4Mq_3\bar{n}^2}{\sqrt{\bar{n}^2}}(L_l - L_{l-1}+8 L_{l-1}^1).\label{A4_static}
\end{align}


\section{Spectral functions $\rho_i$'s}
\label{appE}

Here we list down the explicit expressions for the spectral functions within the limit of small energy transfer. 
\begin{align}
&\rho_1(\omega,q)  = \frac{1}{\pi} {\rm Im}\left(\mathcal{J}_1\Big |_{q_0 = \omega +i\epsilon}\right) \nn\\
&= \frac{1}{\pi D}\Bigl[\Im_{d_1}\left(\Im_{d_3}^2+\Re_{d_3}^2+Q^4-2Q^2\Re_{d_3}\right)\Bigr].\label{rho1_gen}\\
&\rho_2(\omega,q) = \frac{1}{\pi} {\rm Im}\left(\mathcal{J}_2\Big |_{q_0 = \omega +i\epsilon}\right) \nn\\
&= \frac{1}{\pi}\left[\frac{\Im_{d_2}}{\Im_{d_2}^2-(Q^2-\Re_{d_2})^2}\right], \\
&\rho_3(\omega,q) = \frac{1}{\pi} {\rm Im}\left(\mathcal{J}_3\Big |_{q_0 = \omega +i\epsilon}\right) \nn\\
&= \frac{1}{\pi D}\Bigl[\Im_{d_3}\left(\Im_{d_1}^2+\Re_{d_1}^2+Q^4-2Q^2\Re_{d_1} \right)\Bigr],\\
&\rho_4(\omega,q) = \frac{1}{\pi} {\rm Im}\left(\mathcal{J}_4\Big |_{q_0 = \omega +i\epsilon}\right) = 0,
\end{align}
where the denominator $D$ is expressed as
\begin{align}
&D = \Bigl[ \Bigl( -\Im_{d_1}Q^2 - \Im_{d_3}Q^2 + \Im_{d_3}\Re_{d_1} + \Im_{d_1}\Re_{d_3}\Bigr)^2 \nn\\
&+\Bigl(-\Im_{d_1}\Im_{d_3} + (Q^2-\Re_{d_1})(Q^2-\Re_{d_3})\Bigr)^2\Bigr].
\end{align}

Here $\Im_{d_i}$ and $\Re_{d_i}$ respectively depict the imaginary and real parts of $d_i$'s, which we discuss below. The imaginary parts will contribute primarily from $\mathcal{T}_Q$'s and the frequency sum $\Phi_2$ and $\Phi_3$. Respective imaginary parts are given by
\begin{subequations}
\label{imag_parts_di}
\begin{align}
    \Im (\mathcal{T}_Q) \Big|_{q_0\rightarrow \omega + i\epsilon} &= -\frac{\pi\omega}{2q}, \\
    \Im (\Phi_2) \Big|_{q_0\rightarrow \omega + i\epsilon} &=\sum_{s_1,s_2=\pm 1}\frac{1-n_F(s_1E_{k_3})-n_F(s_2E_{r_3})}{4s_1s_2E_{k_3}E_{r_3}}\nn\\
    &~~  \times \pi\delta(\omega-s_1E_{k_3}-s_2E_{r_3}), \\
     \Im (\Phi_3) \Big|_{q_0\rightarrow \omega + i\epsilon} &=\sum_{s_1,s_2=\pm 1}\frac{1-n_F(s_1E_{k_3})-n_F(s_2E_{r_3})}{4s_2E_{r_3}}\nn\\
    &~~  \times \pi\delta(\omega-s_1E_{k_3}-s_2E_{r_3}),
\end{align}
\end{subequations}
where we have used the identity : 
\begin{align}
    \Im \left(\frac{1}{x+i\epsilon} \right) = -\pi\delta(x).
\end{align}

So, the imaginary parts of $d_i$'s can be obtained by replacing $\mathcal{T}_Q$, $\Phi_2$ and $\Phi_3$ in Eqs.~\eqref{coeff_d1_final},\eqref{coeff_d2_final},\eqref{coeff_d3_final} and \eqref{coeff_d4_final} with their respective imaginary parts from Eq.~\eqref{imag_parts_di}. Similarly the real parts of $d_i$ can be obtained by using the respective principal values. 

Let us now specifically discuss the static limit expression for $\left[\frac{1}{\omega}\rho_1^{(s)}(\omega,q)\right]_{\omega\rightarrow 0}$. First we write down the expression for $\Im_{d_1}$.
\begin{align}
    \Im_{d_1} &= \frac{N_c g^2T^2}{3}\frac{\pi\omega}{2q}-\sum_f~\frac{g^2 |q_fB|}{2\pi^2}\sum_{l=0}^\infty (2-\delta_{l,0}) \nn\\
    &\int dk_3 ~E_{k_3}^2~ \Im (\Phi_2)\Big|_{l'=l},
\end{align}
with
\begin{align}
     &\Im (\Phi_2)\Big|_{l'=l} = \frac{\pi\omega}{4E_{k_3}^2}\frac{\partial n_F(E_{k_3})}{\partial E_{k_3}} \delta(\omega-q_3),
\end{align}
where we have restricted ourselves for $\omega >0$ case only. Subsequently looking at the expression of $\rho_1$ in Eq.\eqref{rho1_gen}, we can combine the factor $\frac{1}{\omega}$ with $\Im_{d_1}$ to write down 
\begin{align}
    \frac{1}{\omega}\Im_{d_1}\Big|_{\omega\to 0} &= \frac{N_c g^2T^2}{3}\frac{\pi}{2q}-\sum_f~\frac{g^2 |q_fB|}{4\pi^2}\sum_{l=0}^\infty (2-\delta_{l,0}) \nn\\
    &\times \frac{\pi}{2}\delta(q_3)\int dk_3~ \frac{\partial n_F(E_{k_3})}{\partial E_{k_3}}.
\end{align}
Recalling the expressions for the pure glue part $m_D^g$ and magnetic field dependent correction $\delta m_D$ of the QCD Debye mass, we can readily see that $\frac{1}{\omega}\Im_{d_1}\Big|_{\omega\to 0}$ can be expressed in terms of those Debye mass as 
\begin{align}
    \frac{1}{\omega}\Im_{d_1}\Big|_{\omega\to 0} &= (m_D^g)^2\frac{\pi}{2q}+ \frac{\pi}{2}\delta(q_3) \sum_f~\delta m_{D,f}^2,
\end{align}
where 
\begin{align}
    (m_D^g)^2 &= \frac{N_c g^2T^2}{3}, \nn\\
    \delta m_{D,f}^2 &= -\frac{g^2 |q_fB|}{4\pi^2}\sum_{l=0}^\infty (2-\delta_{l,0})\int dk_3~ \frac{\partial n_F(E_{k_3})}{\partial E_{k_3}},\nn\\
    &= \frac{g^2 |q_fB|}{4\pi^2T}\sum_{l=0}^\infty (2-\delta_{l,0})\int dk_3~ n_F(1-n_F). \nn
\end{align}
Next we proceed to evaluate $\Re_{d_1}$ in the static limit. From Eq.~\eqref{coeff_d1_final}, Eq.~\eqref{Phi2_equal_ll} and using the expressions for the Debye masses mentioned above, one can arrive in a straightforward way to the following expression, 
\begin{align}
    \Re_{d_1}\Big|_{q_0\to 0} = (m_D^g)^2 + \sum_f \delta m_{D,f}^2 = (m_D')^2,
\end{align}
where $m_D'$ now denotes the magnetized medium modified full QCD Debye mass. Looking at the rest of the terms, we do not explicitly write down $\Im_{d_3}$ here, but it can also be seen that at $\omega\to 0$ limit, it vanishes. Eventually the expression of $\left[\frac{1}{\omega}\rho_1^{(s)}(\omega,q)\right]_{\omega\rightarrow 0}$ can be simplified as 
\begin{align}
   \left[\frac{1}{\omega}\rho_1^{(s)}(\omega,q)\right]_{\omega\rightarrow 0} &= \frac{1}{\pi}\frac{ \frac{1}{\omega}\Im_{d_1}\Big|_{\omega\to 0}}{\left(q^2+\Re_{d_1}\Big|_{\omega\to 0}\right)^2}.
   \label{rho1_static}
\end{align}


\end{document}